\documentclass[epj,nopacs]{svjour}
\usepackage{epsfig,array}
\usepackage{multirow}
\usepackage{graphicx}
\def\fun#1#2{\lower3.6pt\vbox{\baselineskip0pt\lineskip.9pt
\ialign{$\mathsurround=0pt#1\hfil##\hfil$\crcr#2\crcr\sim\crcr}}}

\begin{document}
\newcommand{\be}{\begin{eqnarray}}
\newcommand{\ee}{\end{eqnarray}}
\newcommand{\inli}{\int\limits}
\newcommand{\KL}{\rm\Lambda K^+}
\newcommand{\KS}{\rm\Sigma K}
\newcommand{\Ksn}{\rm\Sigma^0 K^+}
\newcommand{\Ksp}{\rm\Sigma^+ K^0}


\title{Decays of Baryon Resonances into \boldmath$\rm\KL$\unboldmath ,
\boldmath$\rm\Ksn$\unboldmath\ and \boldmath$\rm\Ksp$\unboldmath}

\author{
     A.V.~Sarantsev\inst{1,2}
\and V.A.~Nikonov \inst{1,2}
\and A.V.~Anisovich \inst{1,2}
\and E.~Klempt \inst{1}
\and U.~Thoma \inst{1,3}
}
\institute{
Helmholtz--Institut f\"ur Strahlen-- und Kernphysik,
Universit\"at Bonn, Germany
\and Petersburg Nuclear Physics Institute, Gatchina, Russia
\and
Physikalisches Institut, Universit\"at
Gie{\ss}en, Germany
}

\date{\today}

\abstract{
Cross sections, beam asymmetries,  and recoil polarisations for the
reactions $\rm\gamma  p \rightarrow K^+  \Lambda ; \gamma  p
\rightarrow K^+  \Sigma^0$, and $\rm\gamma  p \rightarrow K^0 
\Sigma^+$ have been measured by the SAPHIR, CLAS, and
LEPS collaborations with high statistics and good angular
coverage for centre-of-mass energies between 1.6 and 2.3 GeV.
The combined analysis of these data with data from  $\pi$ and
$\eta$ photoproduction reveals evidence for new baryon
resonances in this energy region.
A new $\rm P_{11}$ state with mass $1840$ MeV and width
$140$ MeV was observed contributing to most of the
fitted reactions. The data demand the
presence of two $\rm D_{13}$ states at 1870 and 2170 MeV.\\
\vspace*{4mm}
}
\titlerunning{\bf Decays of Baryon Resonances into
\boldmath$\rm\KL$\unboldmath ,
\boldmath$\rm\Ksn$\unboldmath\ and \boldmath$\rm\Ksp$\unboldmath
}
\mail{klempt@hiskp.uni-bonn.de}

\maketitle

\section{Introduction}
A quantitative approach to strong interactions at low energies
can not be constructed without detailed information about the
properties of strongly interacting particles. In recent years,
substantial progress had been achieved in understanding the spectrum and
the properties of low mass mesons; the observations of new
mesons consisting of heavy quarks is one of the hottest topics in
hadron physics. However, an understanding of the interaction between
quarks can hardly be reached on the basis of knowing only
quark--antiquark systems. Only baryons can provide
information if pairs of quarks like to cluster into diquarks
or if their excitation spectrum unravels the full richness
of three--particle dynamics. But at present, only the
low--mass baryon resonances are reasonably
well established experimentally~\cite{Krusche:2003ik},
even though the nature of some of these states is
still under discussion. In particlar there is no consensus if the
Roper $\rm N(1440)P_{11}$, the $\rm N(1535)S_{11}$, and the
$\rm\Lambda(1405)S_{01}$ resonances should be interpreted as
excited three--quark states or if they are created by conventional or
chiral meson--baryon interactions
\cite{Meissner:1989nn,Krehl:1999km,Jaffe:2003sg,Kaiser:1995cy,Glozman:1995tb,CaroRamon:1999jf,Steininger:1996xw,Jido:2003cb}.
Above 1.8\,GeV, data become sparse, and
even the density of states is unclear
\cite{Isgur:1995ei}.

Photoproduction of nucleon
resonances in their decay to strange particles offers attractive
possibilities. In the reaction $\rm\gamma p\to \Lambda K^+$, only $\rm
N^*$ and no $\rm\Delta^*$ resonances can contribute. The reactions
$\rm\gamma p\to \Sigma^0 K^+$ and $\rm\gamma p\to \Sigma^+ K^0$
contribute to both the $\rm N^*$ and $\rm\Delta^*$ series, but with
different couplings (due to different Clebsch--Gordan coefficients).
Since baryon resonances have large widths and are often overlapping, a
reduction in the number of partial waves or/and rigid constraints
between different data sets facilitates greatly the task
to identify the leading contributions.

Some of the `missing'
resonances are predicted to couple strongly to
$\rm\Lambda K$ and $\rm\Sigma K$~\cite{Capstick:1998uh}
hence they may contribute significantly
to these particular channels.
Furthermore, $\rm\Lambda$'s reveal their polarisation in their decay, so
that polarisation variables are accessible without use of a polarised
photon beam or a polarised target. The same is true for the $\KS$
reaction through the $\rm \Sigma^0\to \Lambda \gamma$ decay.
And, last not least, high
statistics photoproduction data are now available from
SAPHIR~\cite{Glander:2003jw}, CLAS~\cite{McNabb:2003nf},
and LEPS~\cite{Zegers:2003ux}.
In this letter we present results of a partial wave analysis of these
data, point out differences between the data and common features.
The SAPHIR
data on reaction $\rm\gamma p\to K^0\Sigma^+$~\cite{Lawall:2005np}
were finalized only after completion of most of the fits described
here. They are not included in the systematic evaluation of
errors but only in the final fit. The changes in pole position and
helicity couplings induced by the new SAPHIR data are marginal only.
\par
Data on meson production off nucleons have been interpreted using
different approaches. 
Feuster and Mosel developed a unitary effective Lagrangian model
and described both meson--\ and photon--induced
reactions on the nucleon. Data involving known baryon resonances of spin
$J\leq 3/2$ and later $J\leq 5/2$ were fitted and  $\rm\gamma N$, $\rm\pi N$,
$\rm\pi\pi N$, $\rm\eta N$ and $\rm K \Lambda$ partial widths were extracted
\cite{Feuster:1997pq,Feuster:1998cj,Penner:2002ma,Penner:2002md,Shklyar:2004dy}.
The data have also been interpreted by Regge--model
calculations \cite{Guidal:2003qs}
using only $\rm K$ and $\rm K^*$ exchanges and no $s$--channel resonances.
Only the gross features of the data were reproduced.
Two models based on similar effective Lagrangian
approaches \cite{maid,Mart:1999ed,Janssen:2001pe} gave the correct
order of magnitude of the total
cross section but failed to reproduce differential cross sections.
A structure near 1.9 GeV was
interpreted~\cite{Mart:1999ed} as evidence for a `missing' resonance
at this mass.
Quantum numbers $\rm D_{13}$ were tentatively assigned to
the structure which seemed consistent with the angular distribution and
quark model prediction \cite{Capstick:1998uh}.
Including data on $\rm K^0\Sigma^+$ required
adding the $\rm N(1720)P_{13}$ resonance~\cite{Mart:2000jv}.
In a more recent analysis~\cite{Mart:2003ty}, the structure finds
a more complex interpretation, and the $\rm D_{13}$ partial wave
is found to be resonant at 1740\,MeV and at a higher ill--defined
mass. Janssen et al.~\cite{Janssen:2002jg} do not see the need for 
introducing a $\rm N(1895)D_{13}$ resonance.
Other groups~\cite{Bennhold:2000id,Julia-Diaz:2005qj}
find evidence for a third
 $\rm S_{11}$ resonance which has been suggested to explain an anomaly
in the $\eta$ photoproduction cross section~\cite{Rebreyend:2000se}.
Usov and Scholten~\cite{Usov:2005wy}
 use a coupled channel frame derived from an
effective Lagrangian to fit CLAS and SAPHIR data on $\KL$ and $\KS$.
Eight $\rm N^*$ and 3 $\rm\Delta^*$ resonances are introduced, among them
two  $ \rm P_{11}$ resonances at 1520 and 1850\,MeV and a
$\rm\Delta(1855)P_{33}$.
The CLAS collaboration concluded that interference  between several
resonant states must be important in this mass range, rather than a single
well--separated resonance.

\begin{table}[b!]
\vspace{-0.6cm}
\caption{Data used in the partial wave analysis,
$\chi^2$ contributions and fitting weights.}
\renewcommand{\arraystretch}{1.3}
\begin{center}
\begin{tabular}{lccccr}
\hline\hline
 Observable & $N_{\rm data}$   & $\chi^2$ 
&\hspace*{-2mm}$\chi^2/N_{\rm data}\hspace*{-2mm}$   & Weight&
 Ref. \\
\hline \hline
$\rm\sigma(\gamma p \rightarrow \Lambda K^+)$
 &  720 &   804  &  1.12 &  4    &
\cite{Glander:2003jw} \\
$\rm\sigma(\gamma p \rightarrow \Lambda K^+)$
 &  770 &  1282  &  1.67 &  2    &
\cite{McNabb:2003nf} \\
$\rm P(\gamma p \rightarrow \Lambda K^+)$
 &  202 &   374  &  1.85 &  1    &
\cite{McNabb:2003nf} \\
$\Sigma(\rm\gamma p \rightarrow \Lambda K^+)$
 &   45 &    62  &  1.42 & 15    &
\cite{Zegers:2003ux} \\
\hline
$\rm\sigma(\gamma p \rightarrow \Sigma^0 K^+)$
 &  660 &   834  &  1.27 &  1    &
\cite{Glander:2003jw} \\
$\rm\sigma(\gamma p \rightarrow \Sigma^0 K^+)$
 &  782 &  2446  &  3.13 &  1    &
\cite{McNabb:2003nf} \\
$\rm P(\gamma p \rightarrow \Sigma^0 K^+)$
 &   95 &   166  &  1.76 &  1    &
\cite{McNabb:2003nf} \\
$\Sigma(\rm\gamma p \rightarrow \Sigma^0 K^+)$
 &   45 &    20  &  0.46 & 35    &
\cite{Zegers:2003ux} \\
\hline
$\rm\sigma(\gamma p \rightarrow \Sigma^+ K^0)$
 &   48 &   104  &  2.20 &  2    &
\cite{McNabb:2003nf} \\
$\rm\sigma(\gamma p \rightarrow \Sigma^+ K^0)$
 &  120 &   109  &  0.91 &  5    &
\cite{Lawall:2005np} \\
\hline
$\rm\sigma(\gamma p \rightarrow p\pi^0)$
 & 1106 &  1654  &  1.50 &  8    &
\cite{Bartholomy:04} \\
$\rm\sigma(\gamma p \rightarrow p\pi^0)$
 &  861 &  2354  &  2.74 &  3.5  &
 \cite{GRAAL1} \\
$\Sigma(\rm\gamma p \rightarrow p\pi^0)$
 &  469 &  1606  &  3.43 &  2    &
 \cite{GRAAL1}\\
$\Sigma(\rm\gamma p \rightarrow p\pi^0)$
 &  593 &  1702  &  2.87 &  2    &
 \cite{SAID1}\\
$\rm\sigma(\gamma p \rightarrow n\pi^+)$
 & 1583 &  4524  &  2.86 &  1    &
\cite{SAID2} \\
\hline
$\rm\sigma(\gamma p \rightarrow p\eta)$
 &  667 &   608  &  0.91 & 35    &
\cite{Crede:04} \\
$\rm\sigma(\gamma p \rightarrow p\eta)$
 &  100 &   158  &  1.60 &  7    &
 \cite{Krusche:nv}\\
$\Sigma(\rm\gamma p \rightarrow p\eta)$
 &   51 &   114  &  2.27 & 10    &
 \cite{GRAAL2}\\
$\Sigma(\rm\gamma p \rightarrow p\eta)$
 &  100 &   174  &  1.75 & 10    &
 \cite{GRAAL1}\\
\hline
\hline
\end{tabular}
\end{center}
\label{chi}
\renewcommand{\arraystretch}{1.0}
\vspace{-0.3cm}
\end{table}
\begin{figure}[b!]
\begin{center}
\vspace{-0.6cm}
\epsfig{file=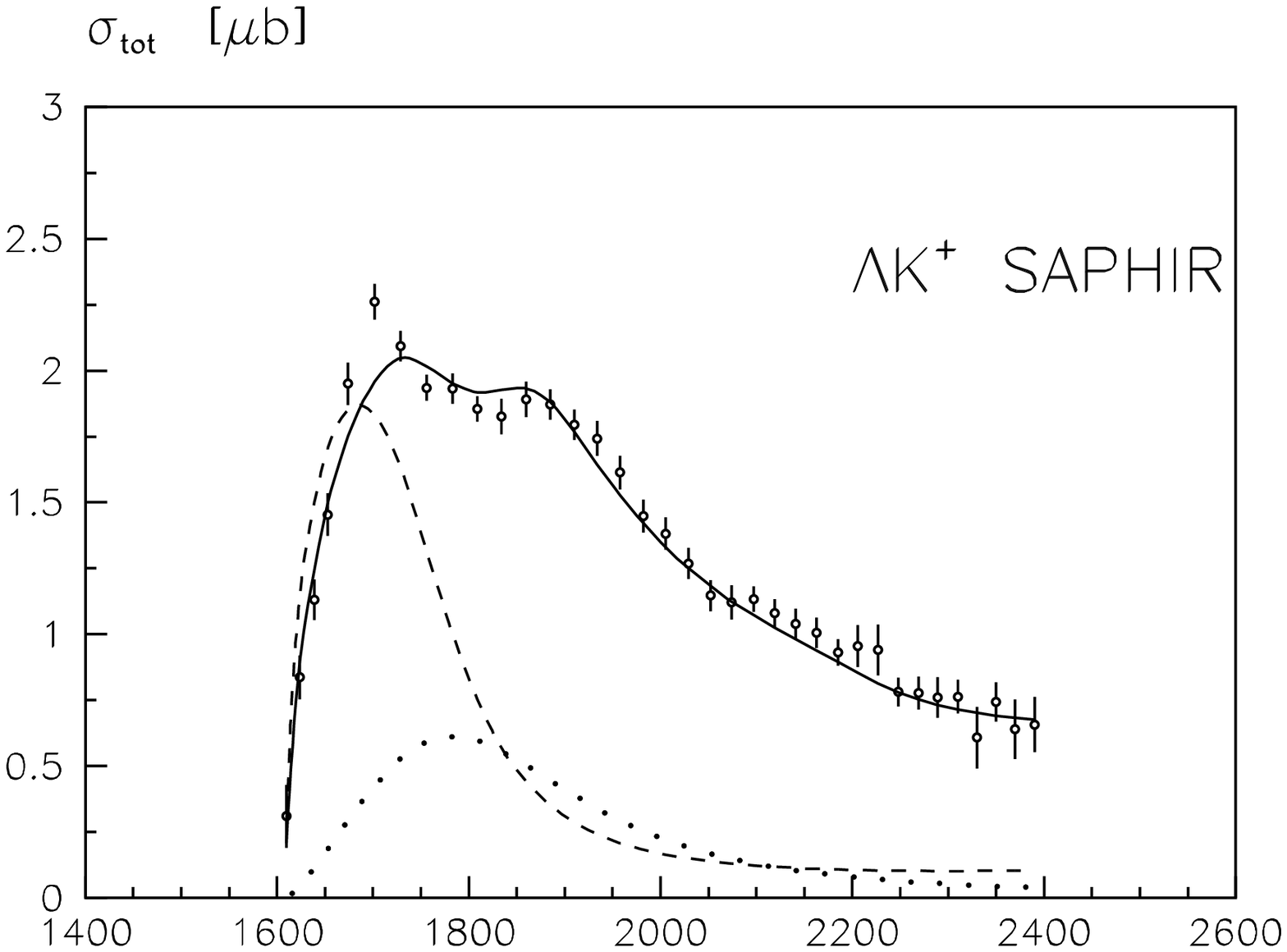,width=0.42\textwidth}\\
\vspace{-1.cm}
\epsfig{file=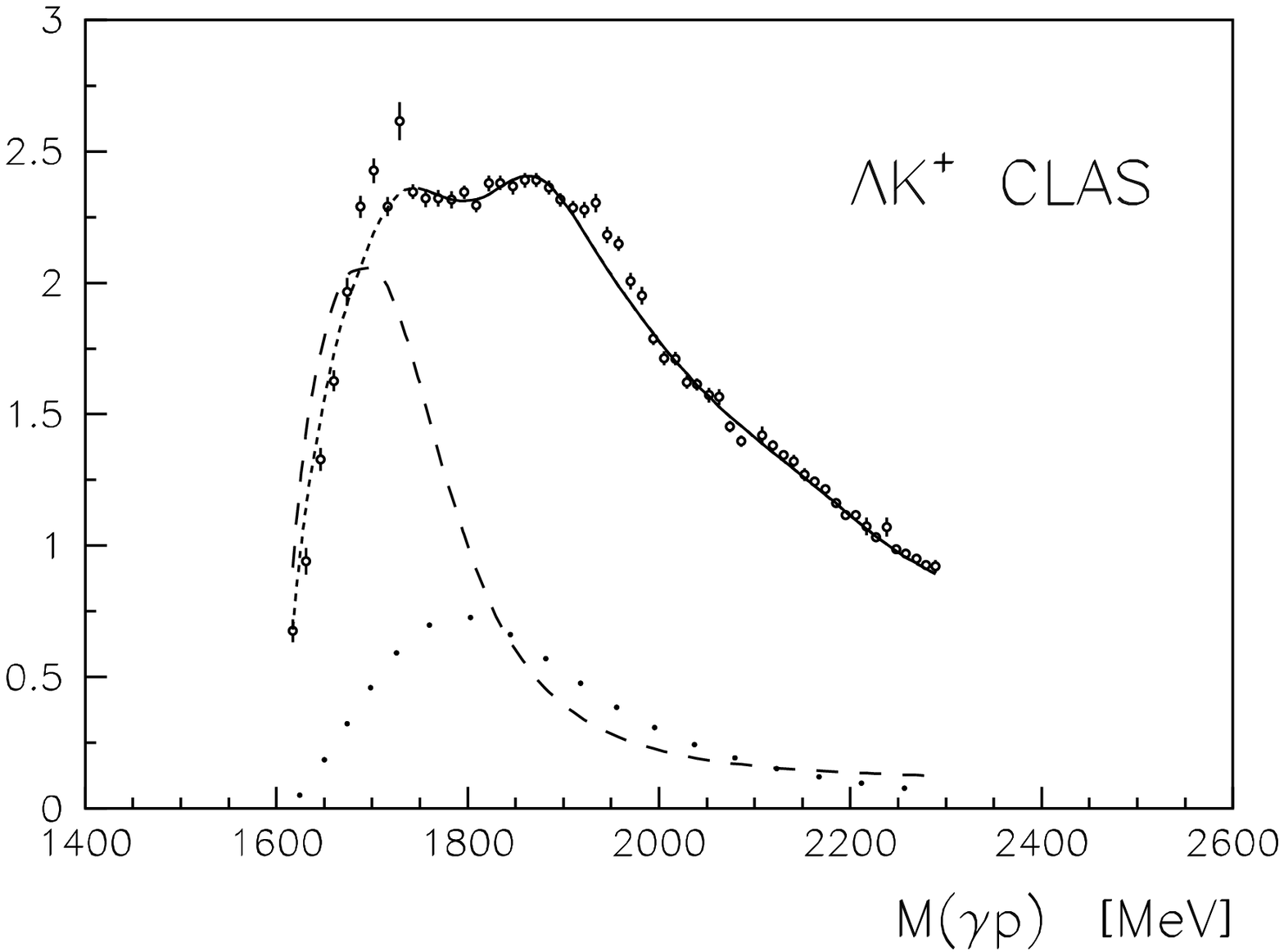,width=0.42\textwidth}
\end{center}
\vspace{-0.1cm}
\caption{\label{fig:kltot}
The total cross sections as a function of $W=\rm M_{\gamma p}$ for
for $\KL$ photoproduction
measured by SAPHIR~\cite{Glander:2003jw}
and CLAS~\cite{McNabb:2003nf}. The solid
curves are results of our fit.
Dashed lines show the contribution from
$\rm S_{11}$ $K$--matrix amplitude, dotted lines show the contribution from
$\rm P_{13}(1720)$. The prediction of the total cross section in the region
where data were not used for the fit is shown as short--dashed line.
The discrepancy at $\rm M_{\gamma p}\sim 1.7$\,GeV is discussed in section
\protect\ref{klambda}.
}
\vspace{0.2cm}
\centerline{\epsfig{file=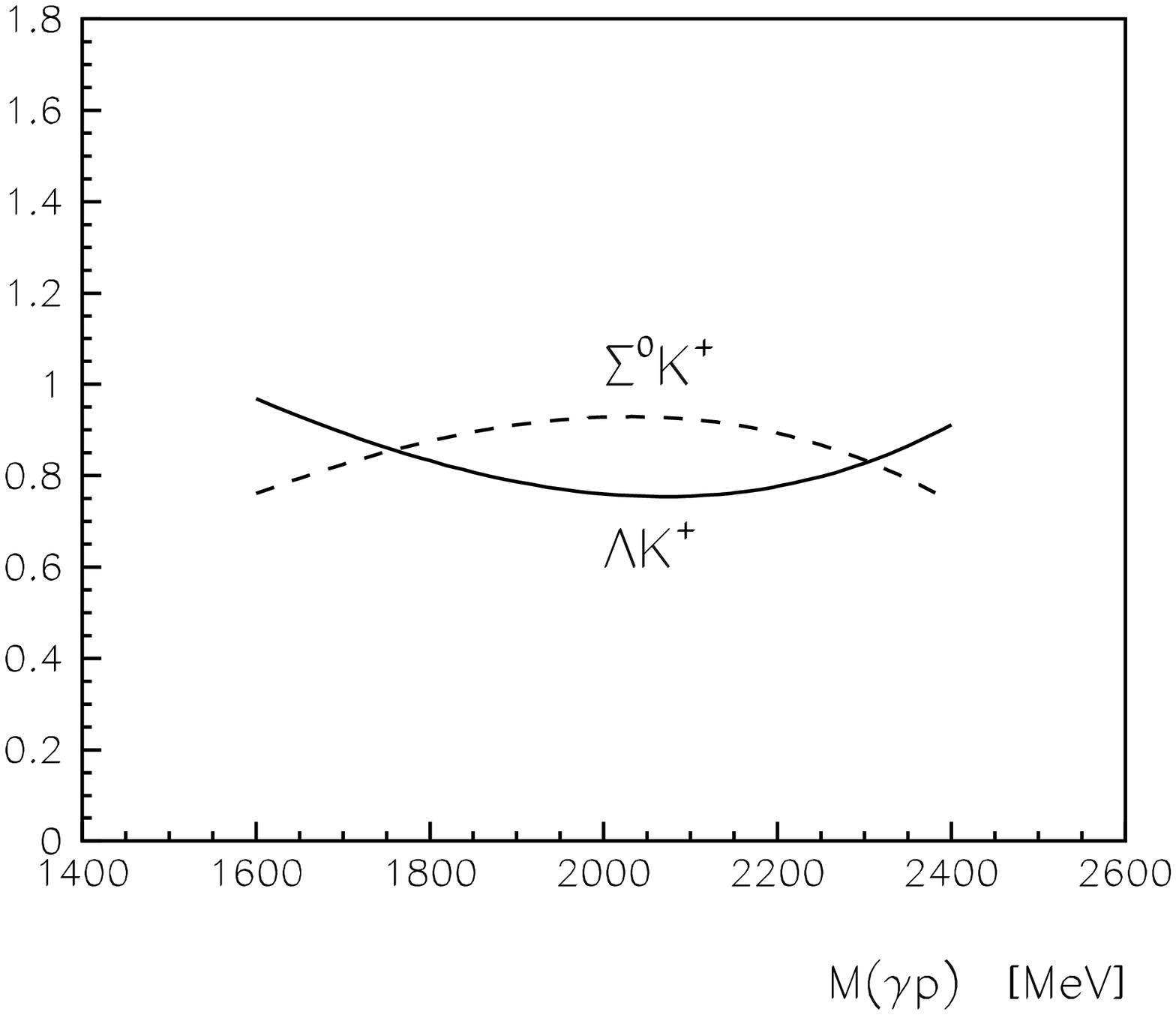,width=0.40\textwidth,
height=0.30\textwidth}}
\vspace{-0.3cm}
\caption{\label{fig:klksren}
Energy dependent normalisation factor, the solid line is for
the $\KL$ data and the dashed line is for the $\KS$ data.
The fit result was multiplied by the factor
before it was compared with the SAPHIR data.}
\vspace{-0.05cm}
\end{figure}
The partial wave analysis presented here is based on the operator expansion
method described in detail in~\cite{Anisovich:2004zz}.
The method is very convenient to describe $s$--channel resonances,
to calculate contributions from triangle and box
diagrams and to project $t$-- and $u$--channel exchange amplitudes
into $s$--channel partial waves.
The data on $\KL$ and $\KS$ were fitted jointly with data
on $\pi^0$ and $\eta$ photoproduction (see Table~\ref{chi}).
Included were the differential cross section for $\pi^0$ and $\eta$
production from CB--ELSA~
\cite{Bartholomy:04,Crede:04}, the Mainz--TAPS
data \cite{Krusche:nv} on $\eta$ photoproduction, cross sections
for $\pi^0$ and $\eta$ photoproduction from GRAAL and beam asymmetry
measurements ~\cite{GRAAL1,SAID1,GRAAL2}, and
data on $\rm\gamma p\rightarrow n\pi^+$~\cite{SAID2}.
The different data sets enter the fits with weights which are listed
in the fifth column of Table~\ref{chi}.
The fits minimise a pseudo--chisquare function
\be
\chi^2_{\rm tot}=\frac{\sum w_i\chi^2_i}{\sum w_i\,N_i}\,\sum N_i\;.
\label{chi_tot}
\ee
where the $N_i$ are given as $N_{\rm data}$ (per channel) 
in the second 
and the weights in the last column  of Table~\ref{chi}.

The partial wave
solutions based on $\pi^0$ and $\eta$ photoproduction
data only were presented earlier in the
two letter publications~\cite{Bartholomy:04,Crede:04}. Aspects of the
new fits related to the $\pi^0$ and $\eta$ photoproduction  are
presented in the preceding
paper~\cite{anis}.
In this paper, those results are discussed which pertain to
final states with open strangeness.

\section{Photoproduction of open strangeness}

\subsection{\label{klambda}
\boldmath$\KL$\unboldmath photoproduction}

The CLAS data cover the mass ($\sqrt s$) range from
the $\rm K\Lambda$ threshold to
2.4\,GeV. The differential angular distributions
are given in 56 bins
about 13 MeV wide at low energies and
10\,MeV wide at high energies. The SAPHIR collaboration showed
36 angular distributions from
threshold to 2.4\,GeV in about 20\,MeV wide mass bins.
The CLAS and SAPHIR data are complemented by
the coincident observation of the $\rm\Lambda$ recoil polarisation.
The LEPS collaboration at SPring-8 published
beam asymmetry measurements in 9 mass bins covering
the same mass range.

\begin{figure}[b!]
\vspace*{-6mm}
\centerline{\epsfig{file=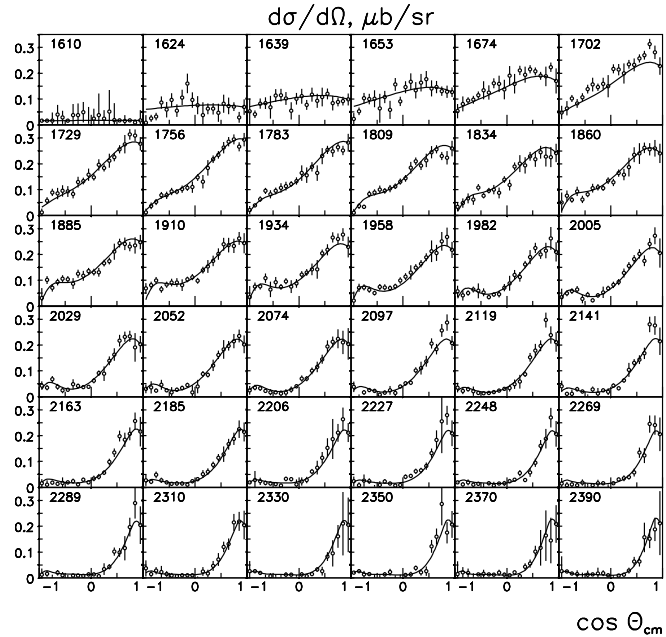,width=0.49\textwidth}}
\caption{The angular distributions as a function of $W$ for
$\rm \gamma p\to\KL$, SAPHIR data.}
\label{fig:klbonn}
\end{figure}

The experimental total cross section was determined
by summation of the experimental differential cross sections and, in
regions where no data exist, predicted values determined by the fit.
For some very forward SAPHIR data points, fit values were taken as well.

The SAPHIR and CLAS $\rm\gamma p\to \KL$ total cross sections show a
very narrow enhancement in the 1700 MeV region. It can easily be fitted
by giving the $\rm N(1650)S_{11}$ resonance a narrow ($\sim 30$\,MeV)
width or by introducing a new $\rm S_{11}$ or $\rm P_{11}$ resonance
(where the former resonance gives a slightly better description). If
such a state exists, it couples strongly only to the $\KL$ channel and
must have a very exotic nature. While the narrow peaks seem to be
consistent in the SAPHIR and CLAS total cross sections, their origin is
very different. In the SAPHIR data, the peak is connected with a larger
differential cross section (compared to the fit) over a broad angular
range; in the CLAS data, the peak originates from  just two points in
the very forward region. If these points are excluded, the CLAS
total cross section is even smaller than given by the fit. Of course, a
new narrow state should not be claimed on this basis; new data are
needed to resolve this discrepancy and the narrow structure at
1700\,MeV is disregarded here.

\begin{figure}[b!]
\vspace*{-6mm}
\centerline{\epsfig{file=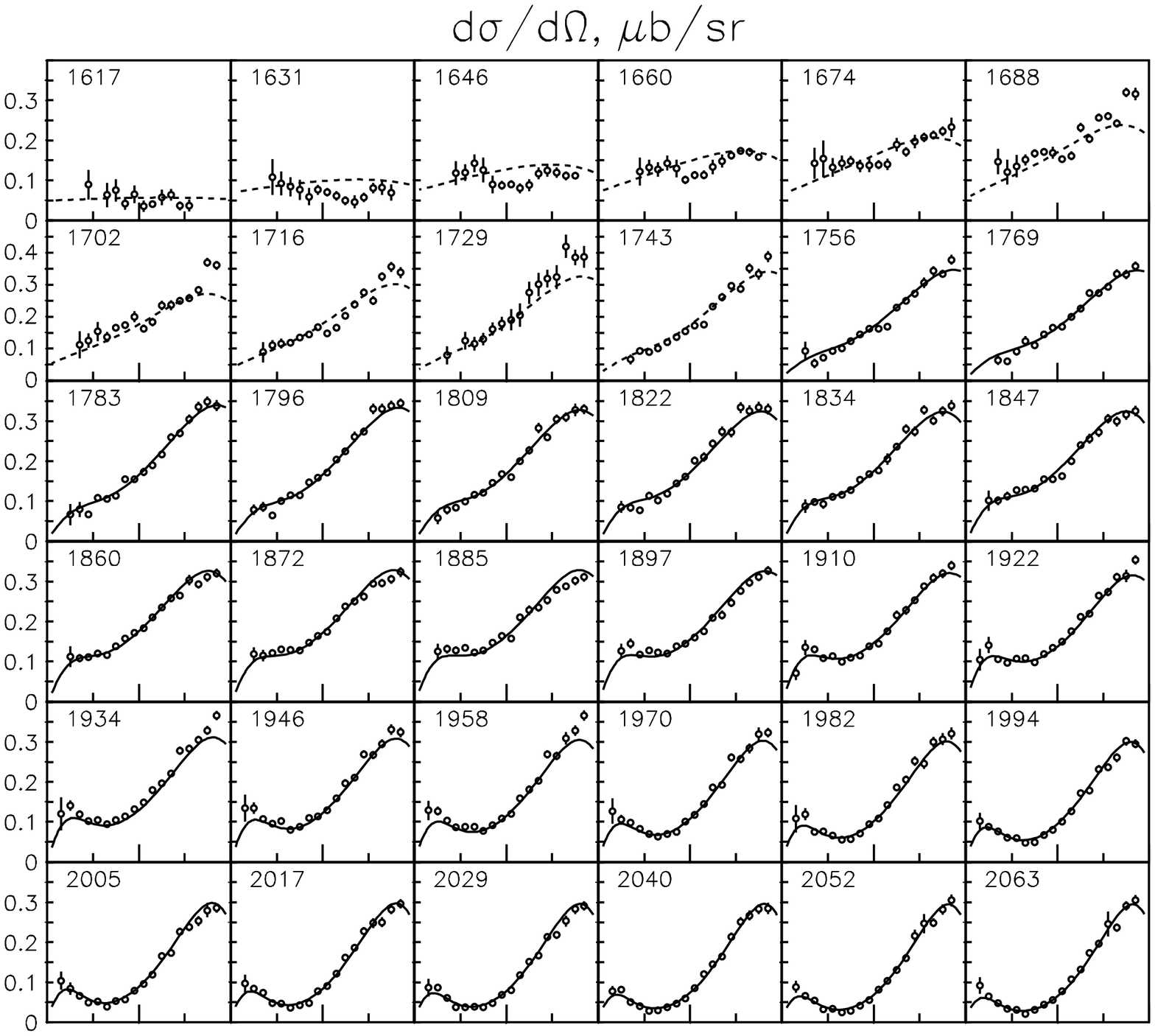,width=0.49\textwidth}}
\vspace*{-16mm}
\centerline{\epsfig{file=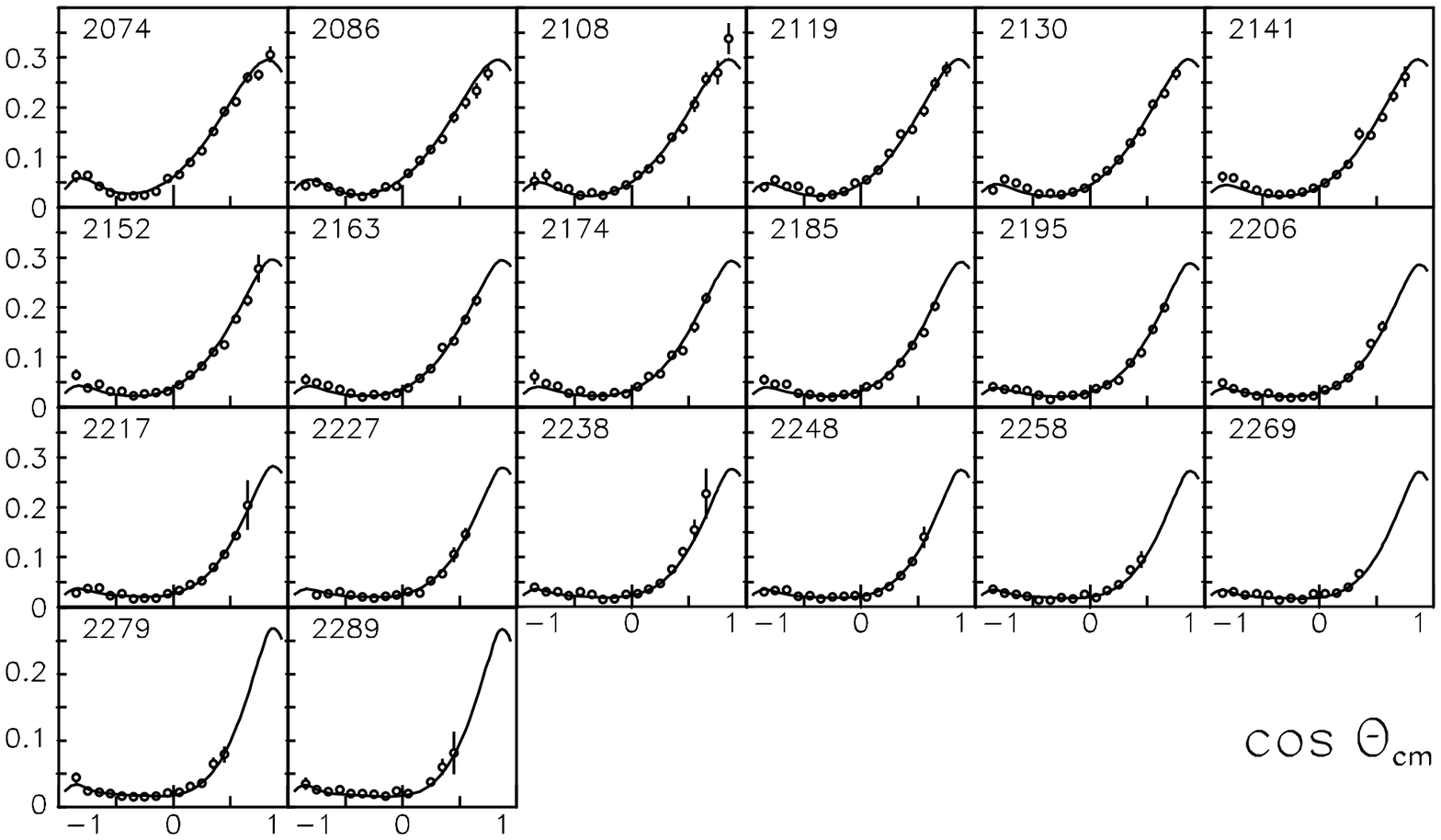,width=0.49\textwidth}}
\vspace*{-30mm}
\caption{\label{fig:klclas}
The angular distributions as a function of $W$ 
for $\rm\gamma p\to\KL$, CLAS
data. The prediction for the region where data were
not used in the fit is shown as dashed lines.}
\end{figure}

The Figs.~\ref{fig:klbonn} and \ref{fig:klclas} show the differential
cross sections obtained by SAPHIR and CLAS and the results of the best
fit described below. The agreement is rather good in both
cases: most discrepancies between the two experiments can obviously
be ascribed to an overall energy--dependent normalisation error.

\begin{figure}[t!]
\vspace{0.55cm}
\centerline{\epsfig{file=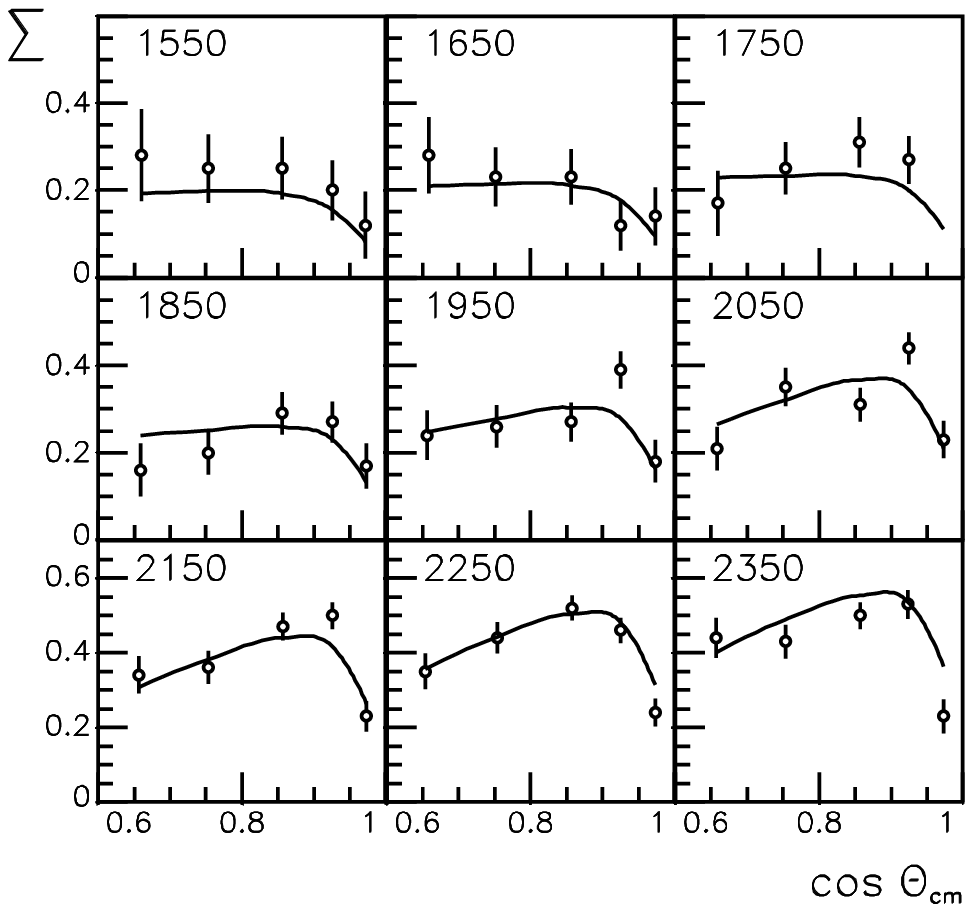,width=0.44\textwidth
,height=0.44\textwidth
}}
\caption{\label{fig:klspri}
The beam polarisation asymmetries as a function of $W$ for
$\rm \gamma p\to K^{+}\Lambda$~\cite{Zegers:2003ux}.
The curves are the result of our fit.}
\vspace{-0.5cm}
\centerline{\epsfig{file=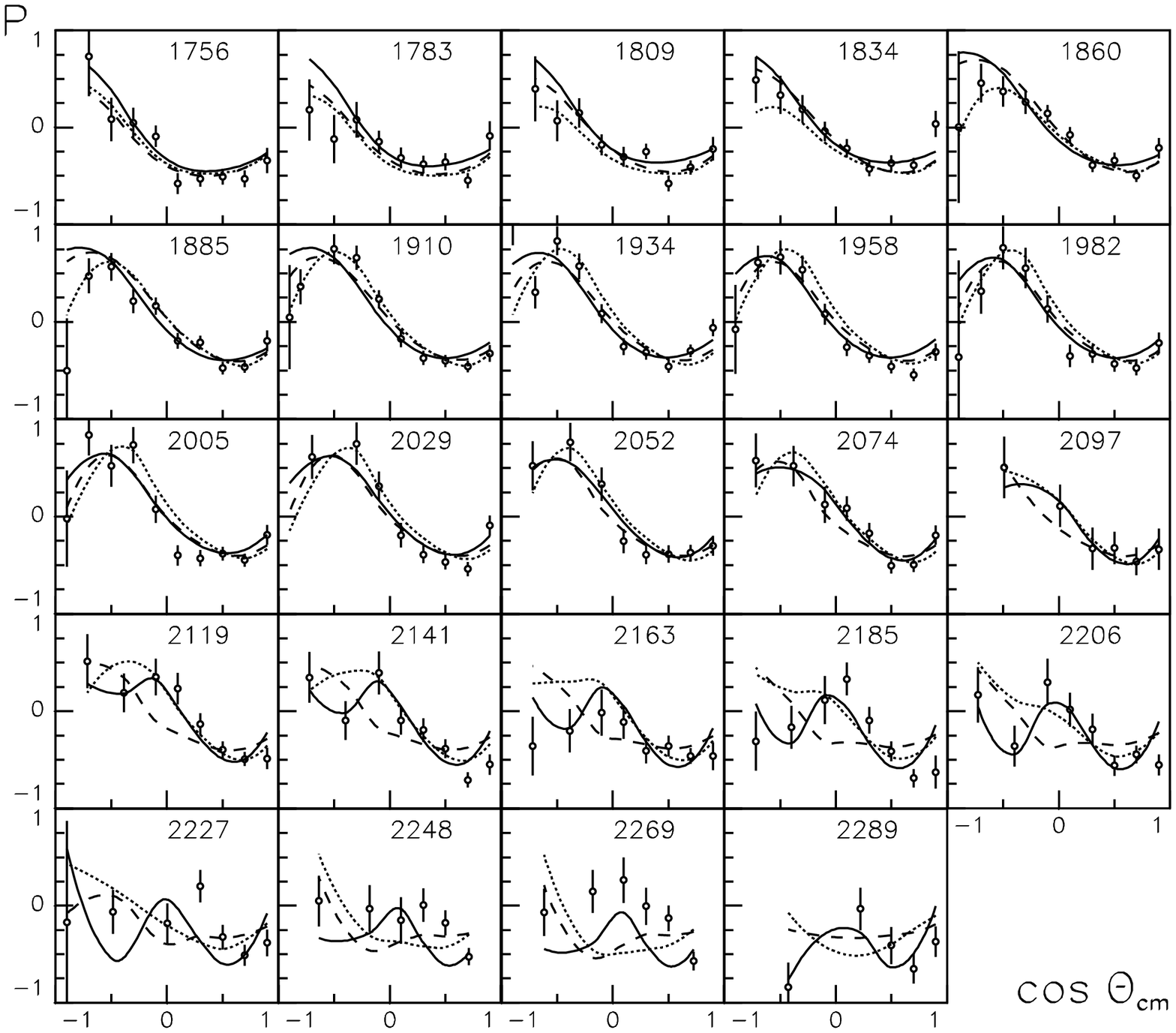,width=0.50\textwidth,height=0.44\textheight}}
\vspace{-0.5cm}
\caption{Recoil polarisation of $\rm \Lambda$ hyperons as a function
of $W$ for the center--of--mass kaon angle 
$\cos(\theta_{K}^{CM})$. The
vertical bars on CLAS data (solid
points) combine statistical and systematic errors.
The solid line presents our fit. The dashed curves represent results
when the $\rm N(2170)D_{13}$ is omitted, for the dotted lines the
$\rm N(1840)P_{11}$ was omitted in the fit.
}
\label{fig:klrecoil}
\vspace{-0.5cm}
\end{figure}

The beam polarisation asymmetries are compared to the fit in
Fig.~\ref{fig:klspri}. For these data, the calculations were made for
narrow bins (with ten times smaller widths) and then averaged to fit
the correspondent experimental data.
The recoil polarisation, obtained
from the weak--decay asymmetry of hyperons, provides further constraints
to the solution. Data and fit, divided into 24 mass bins, are
shown in Fig.~\ref{fig:klrecoil}.

\subsection{$\KS$ photoproduction}

The $\rm\gamma p\to \KS$ data obtained by CLAS and SAPHIR
cover the same energy range as the $\rm\gamma p\to \KL$
data. The data are also complemented by the coincident
observation of the $\rm\Sigma$ recoil polarisation. The beam asymmetry
was measured by LEPS and given in 9 energy intervals covering the
same mass range as the $\KL$ measurements.
The CLAS data for $\Ksp$ photoproduction~\cite{carnahan}
are binned into 6 energy intervals of about 100\,MeV width.
This data is also used in the fits. The calculations were
made for 10\,MeV mass spacings and then averaged
to fit the experimental bins.
\begin{figure}[pb]
\vspace*{-6mm}
\begin{center}
\epsfig{file=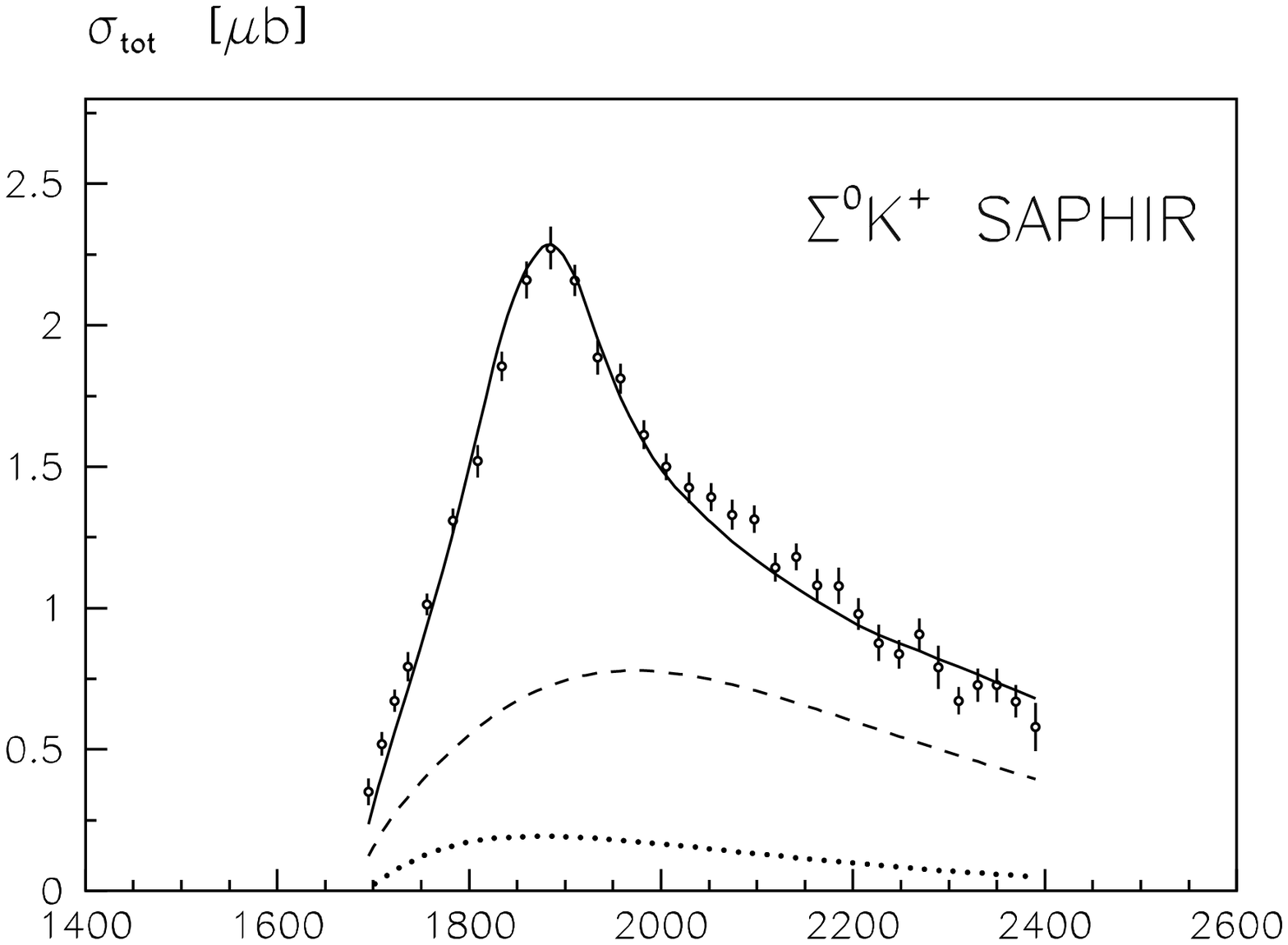,width=0.45\textwidth,
height=0.35\textwidth}\\
\vspace{-1.0cm}
\epsfig{file=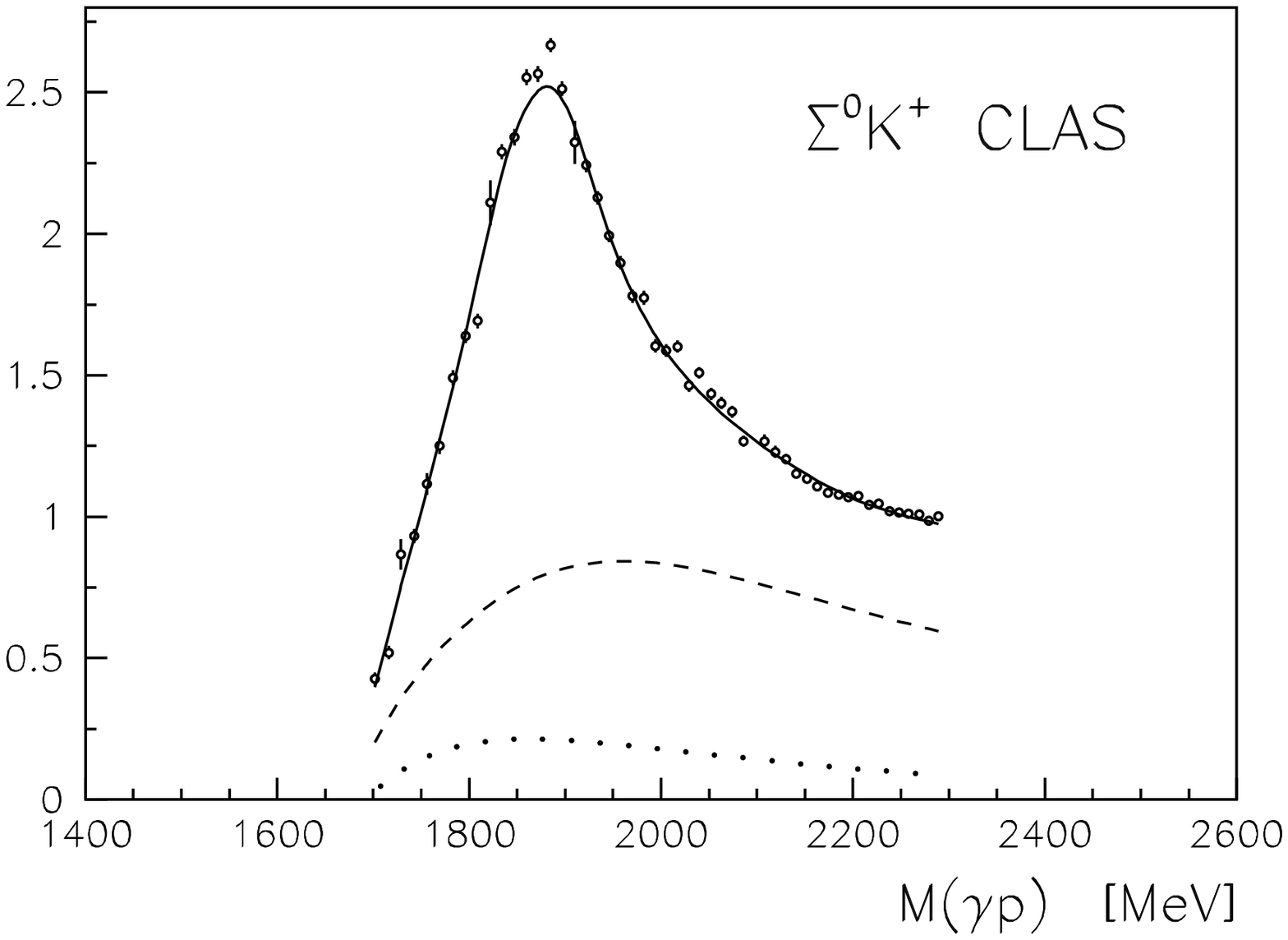,width=0.45\textwidth,
height=0.35\textwidth}
\end{center}
\caption{
Total cross sections for $\Ksn$ photoproduction
measured at {\small SAPHIR}~\cite{Glander:2003jw}
and CLAS~\cite{McNabb:2003nf}. The solid
curves are results of our fit.
The total cross section was determined in the same way as for $\KL$
channel.
Dashed lines show the contribution from $\rm 
K^*$--exchange, dotted lines
the contribution from $\rm D_{33}(1700)$.}
\label{fig:kstot}
\end{figure}

As in the case of the $\KL$ reactions, the total cross section
measurements from SAPHIR and CLAS are not fully compatible. In
Fig.~\ref{fig:kstot} the different height of the total
cross sections as obtained by the two experiments can be seen.
Both data agree only after renormalisation with an energy dependent
function as given in Fig.~\ref{fig:klksren}.

\begin{figure}[pt]
\centerline{\epsfig{file=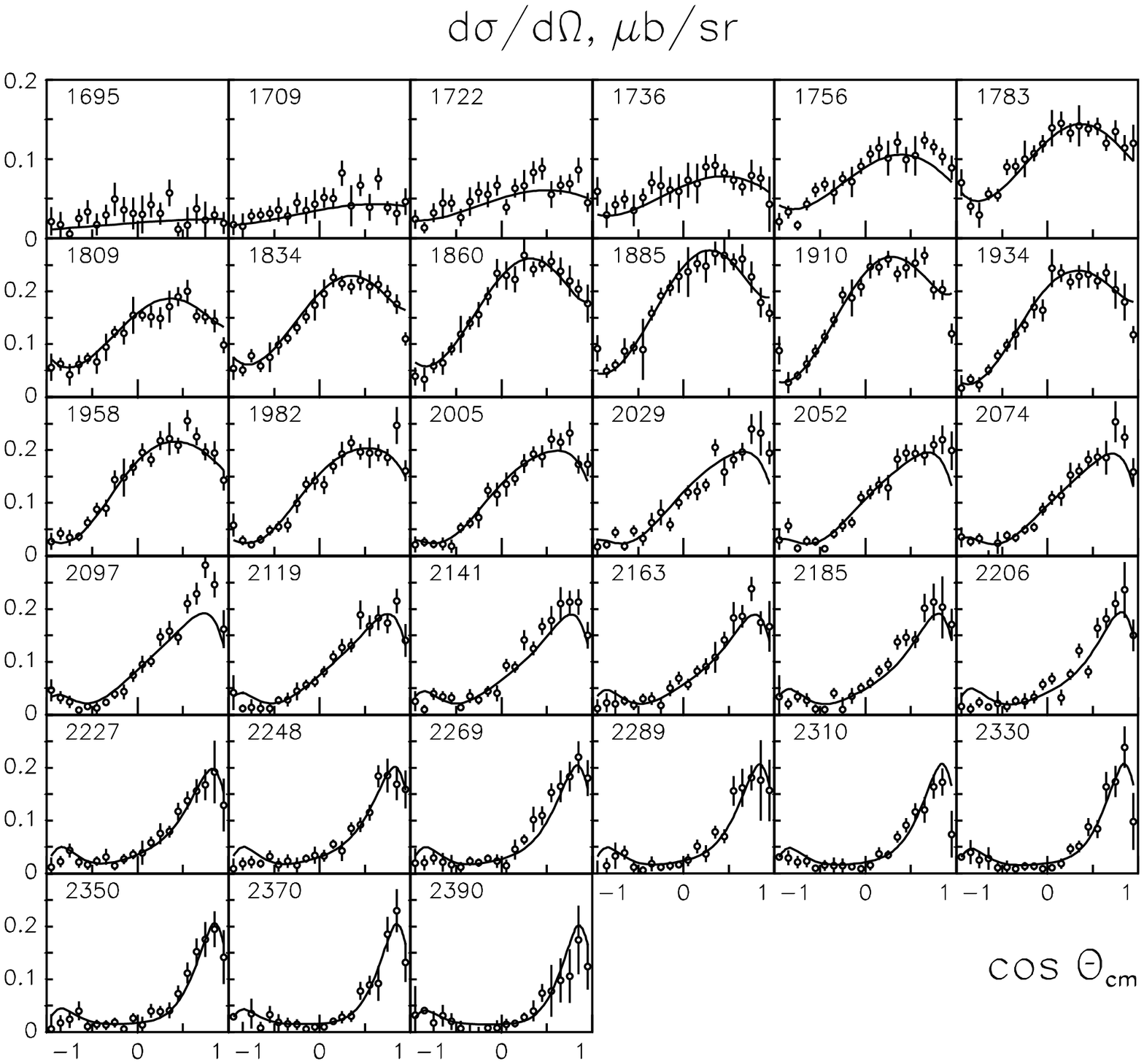,width=0.49\textwidth,height=0.40\textheight}}
\vspace*{-3mm}
\caption{Angular distributions for
$\rm \gamma p\to K^+\Sigma^0$,  SAPHIR data~\protect\cite{Glander:2003jw}.}
\vspace*{3mm}
\label{fig:ksbonn}
\centerline{\epsfig{file=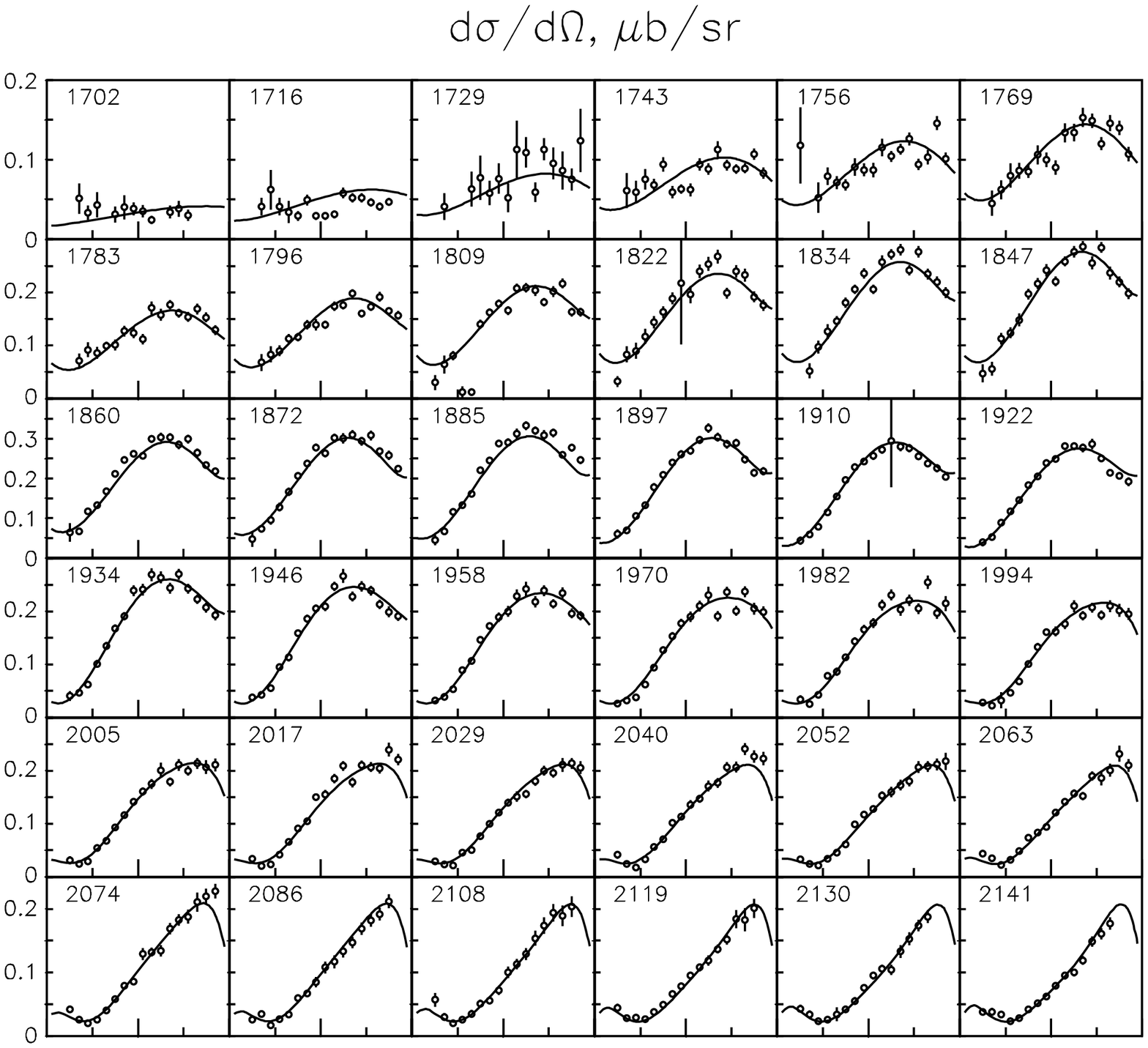,width=0.49\textwidth}}
\vspace{-15.9mm}
\centerline{\epsfig{file=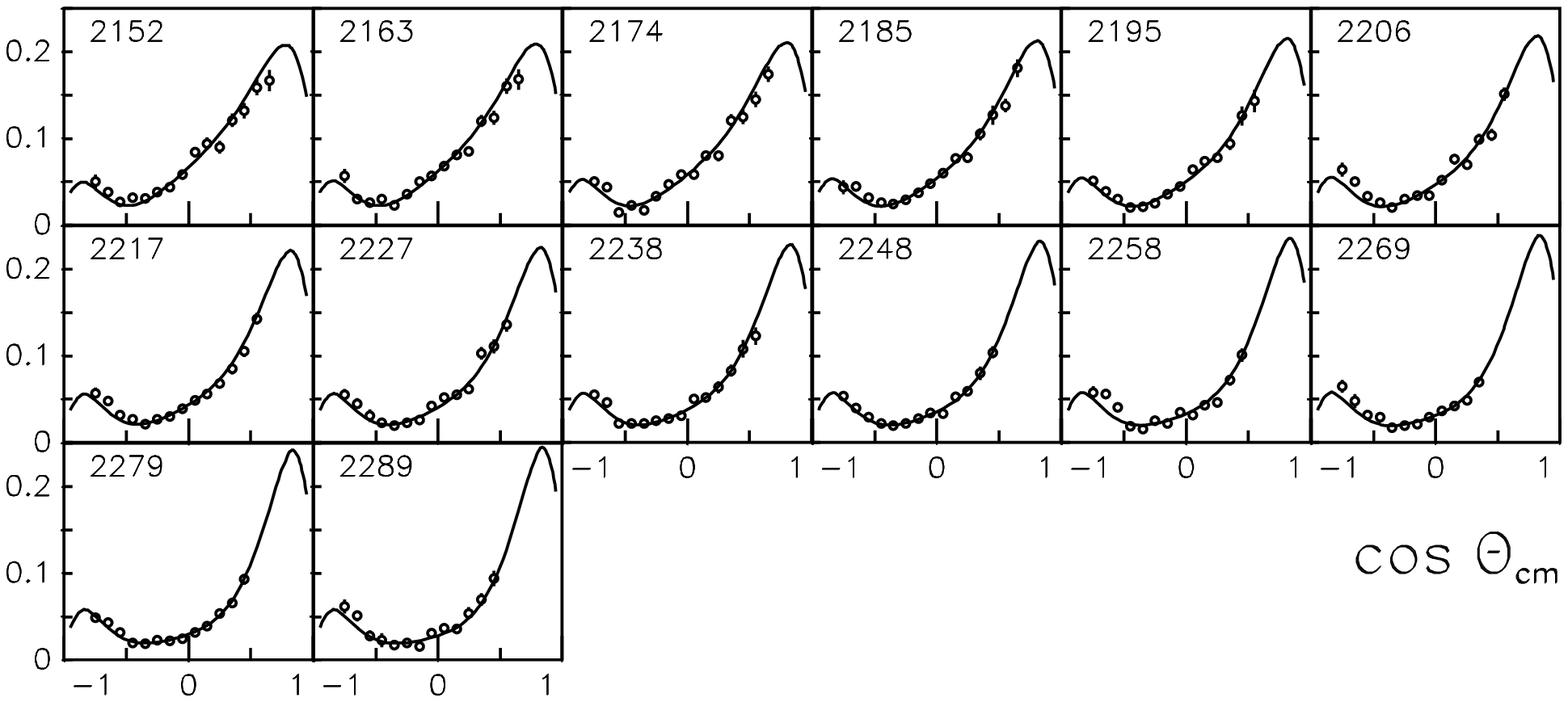,width=0.49\textwidth}}
\vspace{-40mm}
\caption{\label{fig:ksclas}
Angular distributions for $\rm \gamma p\to K^+\Sigma^0$,
CLAS data~\cite{McNabb:2003nf}.}
\end{figure}
\begin{figure}[pt]
\vspace{0.15cm}
\centerline{\epsfig{file=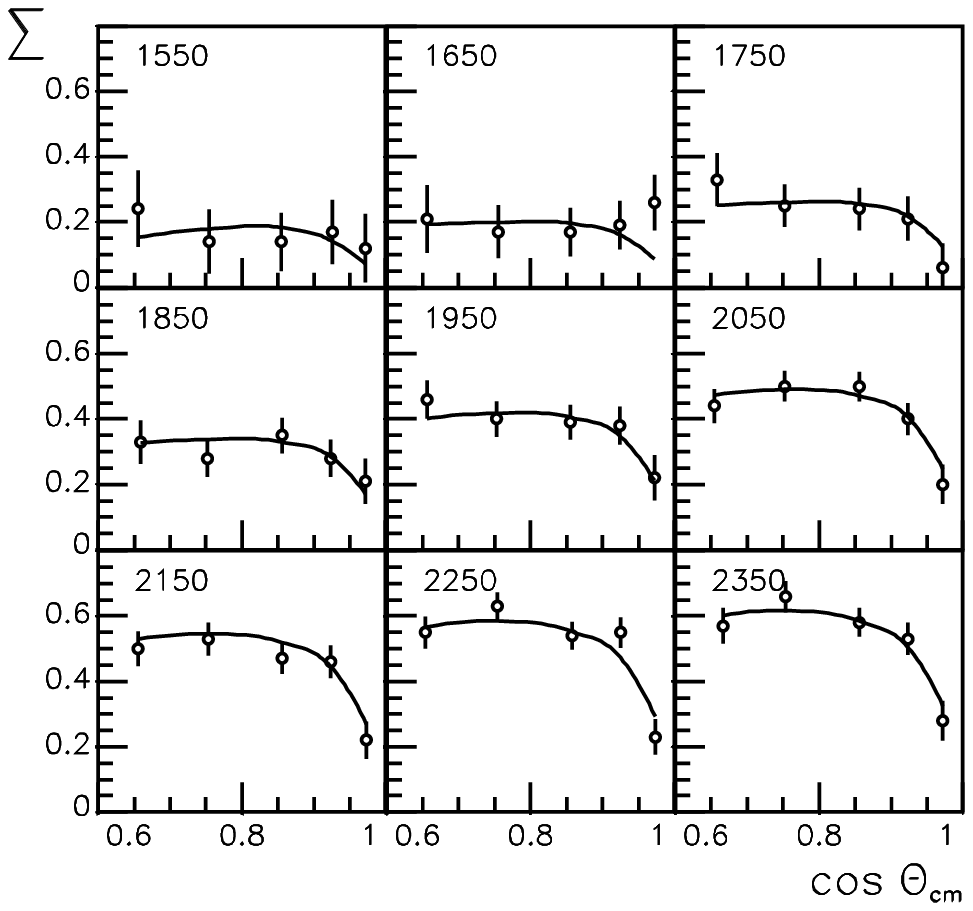,width=0.435\textwidth
}}
\vspace{-0.6cm}
\caption{\label{fig:ksspri}
Beam polarisation asymmetries for
$\rm \gamma p\to K^{+}\Sigma$, LEPS data.}
\vspace{-0.1cm}
\centerline{\epsfig{file=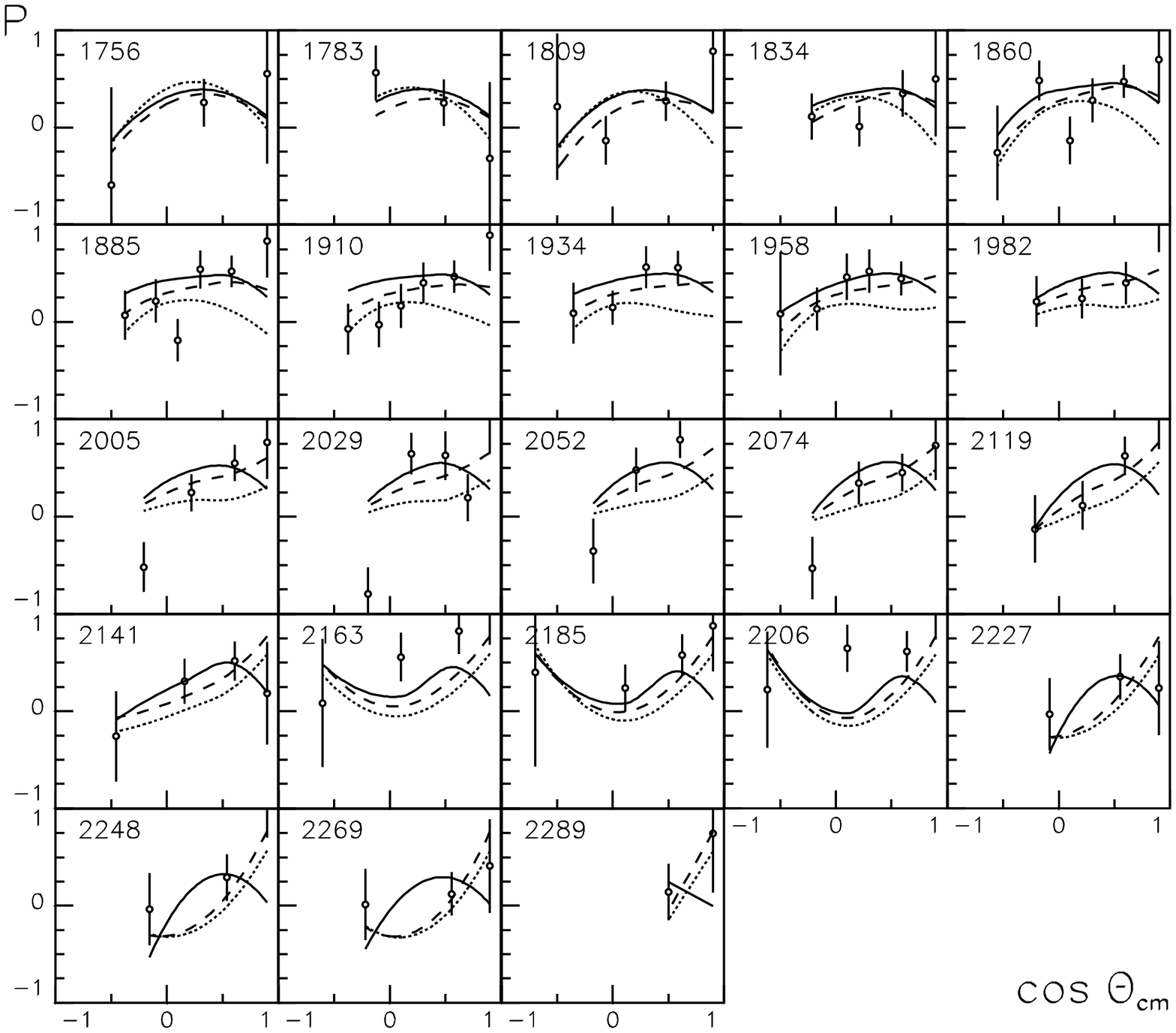,width=0.49\textwidth}}
\vspace{-0.45cm}
\caption{Recoil polarisations for $\rm K^+\Sigma^0$, CLAS data.
The solid curves
are results of the fit. The dashed curves correspond to the
solution without $\rm N(2170)D_{13}$ and the dotted curves to the
solution without $\rm (1840)P_{11}$.}
\label{fig:ksrecoil}
\vspace{-0.3cm}
\end{figure}


The SAPHIR and CLAS differential
cross sections $\Ksn$  and the result of
the best fit are shown in Figs.~\ref{fig:ksbonn} and
\ref{fig:ksclas}. The agreement is rather good in both cases:
as for $\KL$ production, the discrepancy can
be ascribed to an overall energy--dependent
normalisation error. The beam polarisation asymmetries from
LEPS are given in
Fig.~\ref{fig:ksspri} and the
$\rm\Sigma$ recoil polarisation data divided into 23 energy bins
are shown in Fig.~\ref{fig:ksrecoil}.

\begin{figure}[t]
\vspace{-0.35cm}
\epsfig{file=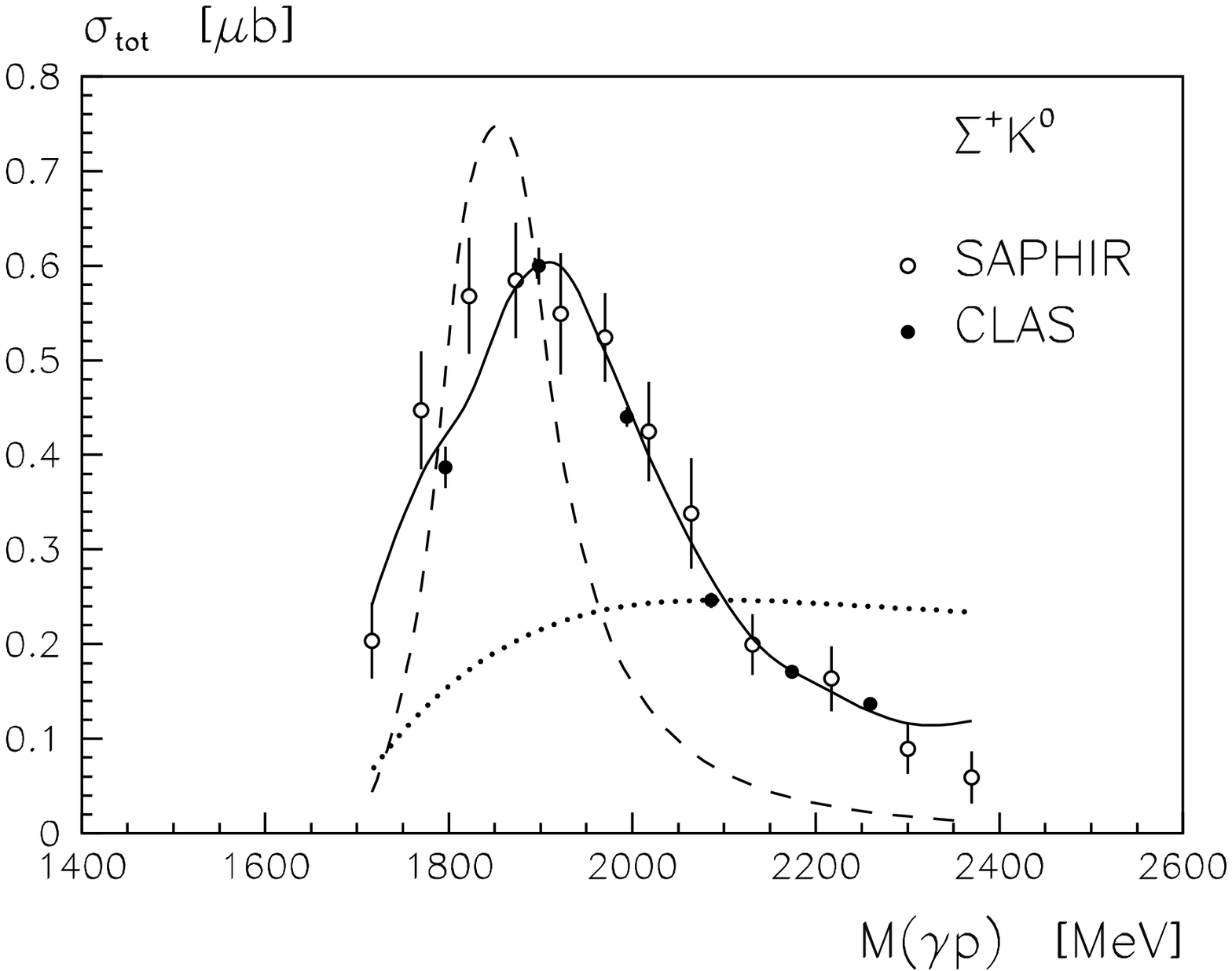,width=0.44\textwidth,
height=0.4\textwidth}
\caption{\label{fig:k0s_tot}
The total cross sections for
$\rm K^0\Sigma^+$ photoproduction.
The solid curve represents our fit.
The SAPHIR cross section is shown as open circles.
For the CLAS data the
experimental total cross section (full circles)
was determined by summation of the
experimental differential cross sections and, in regions where no data
exist, predicted values determined by the fit. Dashed line shows the
contribution from $\rm P_{11}(1840)$, dotted line
shows the contribution from $\Sigma$--exchange.}
\vspace{-0.5cm}
\epsfig{file=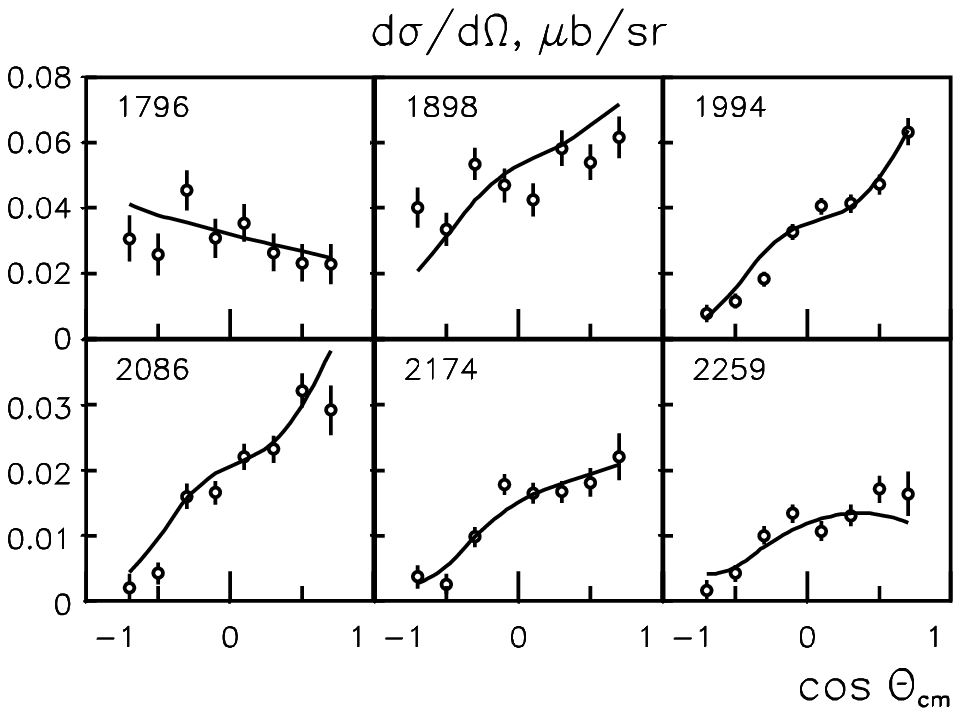,width=0.44\textwidth}
\vspace{-1.5cm}
\caption{\label{fig:k0sc}
Angular distributions for $\rm K^0\Sigma^+$ CLAS data}
\vspace{-0.3cm}
\end{figure}

The total and differential cross sections $\Ksp$  versus our
fit are given in Figs.~\ref{fig:k0s_tot}.
In this case, no normalisation factor was applied. 
\begin{figure}[h!]
\epsfig{file=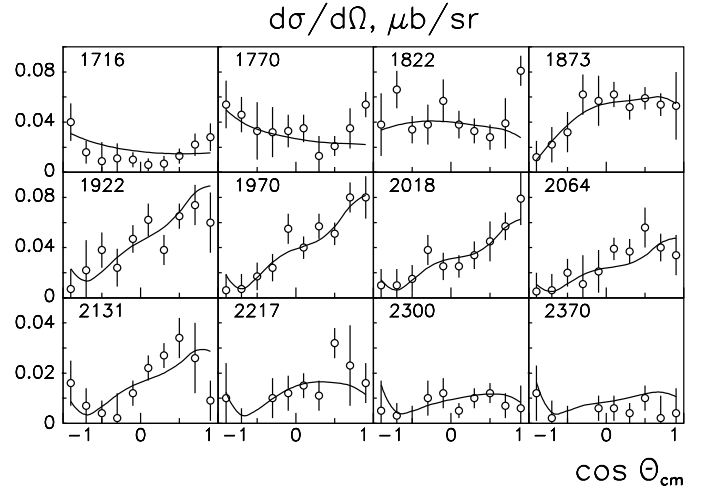,width=0.5\textwidth}
\vspace{-0.8cm}
\caption{\label{fig:k0ss}
Angular distributions for $\rm K^0\Sigma^+$ SAPHIR data}
\vspace{-0.1cm}
\end{figure}

\subsection{\label{comm}The normalisation function}

Finally, a comment seems appropriate concerning the
inconsistency between total cross section obtained by the SAPHIR and
CLAS collaborations. If the problem is connected with an error in the
photon flux normalisation,  the normalisation factor
should not depend on energy. A fit with an energy independent factor
increased $\chi^2$ for $\rm\gamma p\to \KL$ by 16 for the
SAPHIR differential cross sections, and by 40 for the CLAS
differential cross sections. The  recoil polarisation data
were described by a curve with an increase of $\chi^2$
by 20. Such small differences are not seen in the pictures,
and we conclude that the energy dependence does not play any critical
role. The total $\chi^2$ for $\KS$ final states only changed by 20.
A constant normalisation factor was determined to be 
0.80 for $\rm\gamma p\to \KL$  and 0.89 for 
$\rm\gamma p\to \KS$. If all reactions were fitted with
the same factor, it optimised at $0.82\pm 0.03$. This factor is very
close to that obtained for the $\KL$ channel, and the $\chi^2$
changed only by another $20$. However, the calculated
$\KS$ total cross section was then systematically
lower than the SAPHIR data points. In any case, including or excluding
the energy dependence of the normalisation factor does not
change any of the conclusions concerning masses or widths of
baryon resonances.

\section{\label{fit}\vspace*{-2mm}
Fit results}

The fitting weights of the various data sets and their $\chi^2$
contributions are given in Table~1.
The weights of the
$\KL$ and $\KS$ data were selected to obtain a good description of
this data without noticeable deterioration of the fit to $\pi$ and
$\eta$. The solution was carefully checked for stability. The weights
of some channels were changed significantly, the resulting changes
in the fit parameters  were included in the final systematic errors.
For example, increasing the weights from 3 to 35 for the data on beam
asymmetry improved the description of this data, increased slightly
the $\chi^2$ for other data, but lead to very small (maximum 5 MeV)
shifts in resonance positions and/or widths.

\begin{table*}[t!]
\caption{\label{table1}
Masses, widths and coupling constants squared for N$^*$ and $\Delta^*$
resonances. The values are given for the best fit and errors cover
systematical errors. Normalisation condition for Breit--Wigner
resonances is ${\rm\Sigma}\;g_i^2\;=1$. The errors are estimated from
a variety of different solutions.}
\renewcommand{\arraystretch}{1.4}
\begin{center}
\begin{tabular}{lllllll}
\hline\hline
Resonance& {\footnotesize{ M (MeV)}}& {\footnotesize{$\Gamma$ (MeV)}}&
$g^2_{\rm N\pi}$& $g^2_{\rm N\eta}$&$g^2_{\rm \Lambda K}$&$g^2_{\rm \Sigma K}$\\
\hline \hline
$\rm N(1440)P_{11}$&$1450\pm 50$&$250\pm 150$   & 1            &-&-&-\\
PDG          &$1440^{+30}_{-10}$&$350\pm 100$   &              & & & \\
\hline
$\rm N(1520)D_{13}$&$1526\pm 4$&$112\pm 10$     &$0.62\pm 0.06$&$0.04\pm 0.03$&$0.03\pm 0.02$&$0.31\pm 0.09$\ $^{**}$\\
PDG         &$1520^{+10}_{\ -5}$&$120^{+15}_{-10}$&              & & & \\
\hline
$\rm N(1535)S_{11}^*$& $1530\pm 30$&$210\pm 30$ &$0.39\pm 0.10$&$0.95\pm 0.20$&$0.30\pm 0.10$&$0.30\pm 0.10$ \ $^{**}$ \\
PDG           & $1505\pm 10$ &  $170\pm 80$     &              & & & \\
$\rm N(1650)S_{11}^*$&$1705\pm 30$&$220\pm 30$  &$1.10\pm 0.20$ &$0.40\pm 0.10$ & $0.10\pm 0.10$&$0.50\pm 0.15$\\
PDG                    &$1660\pm 20$&$160\pm 10$&              & & & \\
\hline
$\rm N(1675)D_{15}$&$1670\pm 20$&$140\pm 40$    &$0.32\pm 0.15$&$0.04\pm 0.04$&$0.39\pm 0.20$&$0.25\pm 0.20$\\
PDG & $1675^{+10}_{\ -5}$      & $150^{+30}_{-10}$&              & & & \\
\hline
$\rm N(1680)F_{15}$&$1667\pm 6 $&$ 102\pm 15$  &$0.95^{+0.05}_{-0.10}$&$0.00^{+0.05}_{-0.00}$&$0.05^{+0.10}_{-0.05}$&$0.00^{+0.05}_{-0.00}$\\
PDG           &$1680^{+10}_{\ -5}$ & $130\pm 10$ &               & & & \\
\hline
$\rm N(1700)D_{13}$&$1725\pm 15$&$ 100\pm 15$   &$0.29\pm 0.15$&$0.51\pm 0.15$&$0.13\pm 0.10$&$0.07^{+0.12}_{-0.07}$\\
PDG           & $1700\pm 50$ &$100\pm 50$      &               & & & \\
\hline
$\rm N(1720)P_{13}$&$1750\pm 40$&$380\pm 40$   &$0.39\pm 0.10$&$0.43\pm 0.12$&$0.16\pm 0.05$&$0.02\pm 0.02$\\
PDG           &$1720^{+30}_{-70}$& $250\pm 50$ &               & & & \\
\hline
$\rm N(1840)P_{11}$&$1840^{+15}_{-40}$ &$140^{+30}_{-15}$&$0.31\pm 0.10$&$0.09\pm 0.05$ &$0.06\pm 0.03$&$0.54\pm 0.10$\\
 PDG & $1720\pm 30$& $100^{+150}_{\ -50}$ &  \\
\hline
$\rm N(1870)D_{13}$&$1875\pm 25$&$ 80\pm 20$&$0.04\pm 0.04$&$0.21\pm 0.10$&$0.03\pm 0.03$&$0.72\pm 0.30$\\
\hline
$\rm N(2000)F_{15}$&$1850\pm 25$&$225\pm  40$   &$0.85\pm 0.20$&$0.07^{+0.11}_{-0.07}$&$0.03^{+0.07}_{-0.03}$&$0.05^{+0.10}_{-0.05}$\\
PDG           & $\sim 2000$&        \\
\hline
$\rm N(2070)D_{15}$&$2060\pm 30$&$340\pm 50$&$0.71\pm 0.10$&$0.26\pm 0.05$&$0.01\pm 0.01$&$0.02\pm 0.02$\\
\hline
$\rm N(2170)D_{13}$&$2166^{+25}_{-50}$&$ 300\pm 65$   &$0.67\pm 0.30$&$0.15\pm 0.05$&$0.16\pm 0.15$&$0.14^{+0.15}_{-0.14}$\\
PDG           & $\sim 2080$&         \\
\hline
$\rm N(2200)P_{13}$&$2200\pm 30$&$190\pm 50$    &$0.08^{+0.12}_{-0.08}$&$0.89^{+0.08}_{-0.15}$&$0.02^{+0.08}_{-0.02}$&$0.01^{+0.08}_{-0.01}$\\
\hline\hline
$\rm\Delta(1232)P_{33}$&$1235\pm 4$&$140\pm 12$ & 1 & & &-\\
                   PDG &$1232\pm 2$&$120\pm5$   &   & & & \\ \hline
$\rm\Delta(1620)S_{31}$&$1635\pm 6 $&$106\pm 12$&$1.00\pm 0.10$    & & &$0.0^{+0.10}_{-0.00}$\\
PDG               &$1620^{+55}_{\ -5}$&$150\pm30$&               & & & \\ \hline
$\rm\Delta(1700)D_{33}$&$1715\pm 20$&$240\pm 35$&$0.78\pm 0.15$&&&$0.22\pm 0.15$\\
PDG              &$1700^{+70}_{-30}$&$300\pm100$&              & & & \\
\hline
$\rm\Delta(1905)F_{35}$&$1870\pm 50$&$370\pm 110$&$0.90\pm 0.10$ & & &$0.10\pm 0.10$\\
PDG        &$1905^{+15}_{-35}$&$350^{+90}_{-70}$&              & & & \\
\hline
$\rm\Delta(1920)P_{33}$&$1996\pm 30$&$380\pm 40$&$0.94^{+0.06}_{-0.10}$ & & &$0.06^{+0.10}_{-0.06}$\\
PDG       &$1920^{+50}_{-20}$&$250^{+100}_{-50}$&              & & & \\
\hline
$\rm\Delta(1940)D_{33}$&$1930\pm 40$&$200\pm 100$&$0.32\pm 0.30$&&&$0.68\pm 0.30$\\
\hline
$\rm\Delta(1950)F_{37}$&$1893\pm 15$&$240\pm 30$&$0.90\pm 0.10$  & & &$0.10\pm 0.10$\\
PDG       &$1950\pm{10}$    & $ 300^{+50}_{-10}$&              & & & \\
\hline
\hline
\end{tabular}\\
\end{center}
\footnotesize{\hspace*{10mm}$^*$ K matrix fit, the position
of the amplitude poles in the complex planes closest to the physical
region.\\
\hspace*{10mm}$^{**}$ See text for discussion of the large values of
$g^2_{\rm \Sigma K}$.
}
\renewcommand{\arraystretch}{1.0}
\vspace*{-4mm}
\end{table*}
\subsection{First fits}

In first fits, all resonances seen in the analysis of $\pi^0$ and
$\eta$ photoproduction \cite{Bartholomy:04,Crede:04}
were introduced in the fit.
Unphysical solutions were obtained in some of the fits, and
couplings of resonances had to be restricted.
The $\KS$ data have a rather pronounced peak
in the 1800 MeV mass region; some solutions described the peak
by a single resonance having a huge coupling to this channel.
This amplitude created large interferences with other
contributions. To avoid this class of solutions, we demanded
that the $\KL$ and $\KS$ couplings should not
exceed the couplings to the $\rm\pi p$ or $\rm\eta p$ channels
by more than a factor 2. In this way, solutions were found
producing an acceptable overall description of the data
with $\chi^2$ values very close to the best fit not having
such restrictions. In the new fits, all resonances had couplings well
within the boundaries. When a coupling constant fell onto a
boundary value, the fit was not sensitive to this coupling.

In some particular mass regions, the fit deviated visibly from the data.
These regions are now discussed in some detail. The discussion will lead
to the final fit and to the results gathered in
Table~\ref{table1}.

To estimate systematic errors, fits were performed
excluding the SAPHIR or CLAS cross sections.
Masses and widths of the resonances of the main solution are
defined more precisely by the CLAS data; masses
and widths changed by less then 4 MeV in the fit with
the SAPHIR data excluded. If the CLAS cross sections were excluded,
a few more significant changes of the resonance positions were found.
All these uncertainties are included in the errors given in
Table~\ref{table1}.
\par
In Table~\ref{widths} we give ratios of partial widths derived from
the couplings given in Table~\ref{table1}. 
The relation between coupling constants and partial widths is given
e.g. in eq.~(11) of the preceding paper~\cite{anis}.

\begin{table}[b!]
\vspace* {-4mm }
\caption{Ratios of partial widths}
\renewcommand{\arraystretch}{1.2}
\begin{center}
\begin{tabular}{llll}
\hline\hline
Resonance&
$\Gamma_{\rm N\eta}/\Gamma_{\rm N\pi}$&
$\Gamma_{\rm \Lambda K}/\Gamma_{\rm N\pi}$&
$\Gamma_{\rm \Sigma K}/\Gamma_{\rm N\pi}$\\
\hline \hline
$\rm N(1520)D_{13}$&1.5 $\cdot 10^{-3}$&0&0\\
\hline
$\rm N(1675)D_{15}$&0.05 &0.05 &0\\
\hline
$\rm N(1680)F_{15}$&1 $\cdot 10^{-3}$&1 $\cdot 10^{-4}$&0\\
\hline
$\rm N(1700)D_{13}$&0.80&0.07&5 $\cdot 10^{-3}$\\
\hline
$\rm N(1720)P_{13}$&0.80&0.20&0.01\\
\hline
$\rm N(1840)P_{11}$&0.25&0.11&0.80\\
\hline
$\rm N(1870)D_{13}$&2.0&0.28&1.6\\
\hline
$\rm N(2000)F_{15}$&0.04&5 $\cdot 10^{-3}$&3 $\cdot 10^{-3}$\\
\hline
$\rm N(2070)D_{15}$&0.30&8 $\cdot 10^{-3}$&0.015\\
\hline
$\rm N(2170)D_{13}$&0.04&0.17&0.14\\
\hline
$\rm N(2200)P_{13}$&2.0&0.18&0.11\\
\hline\hline
$\rm\Delta(1700)D_{33}$&&&2.5 $\cdot 10^{-3}$\\
\hline
$\rm\Delta(1920)P_{33}$&&&0.04\\
\hline
$\rm\Delta(1940)D_{33}$&&&0.75\\
\hline
$\rm\Delta(1950)F_{37}$&&&0.01\\
\hline
\hline\hline
\end{tabular}\\
\end{center}
\renewcommand{\arraystretch}{1.0}
\label{widths}
\vspace* {-4mm }
\end{table}

\subsection{\label{lowsection}
The low--mass region (1500--1750\,MeV)}
The cross section for the reaction $\rm\gamma p\to\KL$ rises steeply
above threshold and reaches a maximum value at about 1720\,MeV,
just 100\,MeV above the $\KL$ threshold. The peak cross section
for $\rm\gamma p\to\KS$ is reached only at about 1900\,MeV. This
behaviour suggests that near--threshold resonances should have
strong couplings to $\KL$. The fit gives, however, stronger
subthreshold
couplings to the $\KS$ channel. These strong subthreshold $\KS$
couplings parameterise a background which interferes near threshold
with $t$-- and $u$--channel exchanges.


The low--mass part
is strongly influenced by the $\rm S_{11}$ partial wave, in particular
by $\rm N(1650)S_{11}$. As mentioned, the $\rm S_{11}$ partial
wave is described by a two--pole four--channel $K$--matrix.
The lower $K$--matrix pole can have its position in a very wide
mass interval,
in some fits it moved down to 1100 MeV. However, there is no visible
change in the fit as long as the $K$--matrix pole is situated between
1200 and 1460 MeV. For $K$--matrix masses above 1400\,MeV,
the fit started to exceed the SAPHIR total cross
section for $\rm\gamma p\to \KL$ at $\sim$ 1850 MeV, but
the CLAS data were described more precisely. The $K$--matrix
pole is obviously defined only with an appreciable error, we quote
$(1440^{+\ 40}_{-180})$\,MeV. There is a strong dependence of the
low mass $K$--matrix pole couplings on the pole position. For example
when the mass of lowest $K$--matrix pole
goes to lower values the $\KS$
coupling increased very significantly providing the same overall
$\rm S_{11}$ contribution to the $\KS$ cross section.

The $T$--matrix poles are studied in the $\sqrt s$ complex plane.
Due to the four--channel nature, the complex plane is split into
8 Riemann sheets, 4 of them are relevant for the discussion.
The pole positions of the $T$--matrix amplitude are given in
Table~\ref{table1} together with squared $K$--matrix couplings.
The analytical continuation of
\be
\rho_{a}(s)=\frac{\sqrt{(s-(m_\mu+m_B)^2))(s-(m_\mu-m_B)^2)}}{s}\;,
\nonumber \\
a=\pi\mbox{$\rm N, \eta N, K \Lambda, K \Sigma$}\qquad
\mu=\pi,\eta,\mbox{$\rm K$}\qquad B=\mbox{$\rm N,\Lambda,\Sigma$}
\phantom{44}.
\label{phv}
\ee
to the lower complex
$s$--plane defines the sheet closest to the physical region
above threshold, for $Re(s)>(m_\mu+m_B)^2$. For points
on this sheet with $Re(s)<(m_\mu+m_B)^2$, the closest physical region
is at the threshold. Let us denote this sheet as $H$.
The sheet defined by the analytical continuation of the
expression
$$
\rho_{a}(s)=i\frac{\sqrt{((m_\mu+m_B)^2)-s)(s-(m_\mu-m_B)^2)}}{s}
\label{phv_L}
$$
is closest to the physical region for $Re(s)<(m_\mu+m_B)^2$; this
sheet is denoted as $L$. The first pole situated on the sheet
$HHLL$ with respect to the $\rm\pi N$, $\rm\eta N$,
$\rm K\Lambda$, and $\rm K\Sigma$
thresholds has a mass $(1534-i\cdot 107)$\,MeV. The correspondent pole
situated on the sheet $HHHL$ is very close in mass,
$(1518-i\cdot 121)$\,MeV. The second pole has a mass
$1710-i\cdot 105$\,MeV and is situated on the sheet $HHHH$. 
The correspondent pole
on the sheet $HHHL$ has a mass of $1703-i\cdot 115\,MeV$. The 
thresholds situated
near poles have only a weak influence on pole positions.

We found that $T$--matrix poles close to the physical region
are very stable when the masses of the $K$--matrix poles are scanned in a
large interval. Only a few minor changes occurred compared to the previous
analysis~\cite{Bartholomy:04,Crede:04} in which a
two channel $K$--matrix was used. The errors given in the
Table~\ref{table1} are defined from a large set of solutions made
under different assumptions; changes of the pole positions
could even be larger than the errors quoted in the Table, though, 
when more channels are included.

To estimate unseen contributions from three--body final states,
fits were performed with a five--channel $K$--matrix
where the fifth channel provided an unknown inelasticity. It was
parametrised as $\rm\pi\pi N$ phase volume. We found negligible
contributions from this channel and poor convergency of the fits.

\subsection{The intermediate mass range (1700--2200\,MeV)}

Small discrepancies were also seen in different distributions
in the 1800--1900 MeV mass region. To resolve these
discrepancies, resonances with different quantum
numbers were added one by one. The most significant improvement
came from a $\rm P_{11}$ state with mass $(1840^{+15}_{-40})$\,MeV and
width $(140^{+35}_{-15})$\,MeV.
\begin{table}[pb]
\caption{
\label{kl_changes}
Changes in $\chi^2$ when one of the new resonances is omitted
or replaced by a resonance with different spin and parity $J^P$.
The changes are given for the $\chi^2_{\rm tot}$ (\ref{chi_tot})
and the $\chi^2$ contributions for individual final states calculated
analogously.}
\begin{footnotesize} \renewcommand{\arraystretch}{1.2}
\begin{center}
\begin{tabular}{c|ccccc}
\hline\hline
Resonance
&\multicolumn{5}{c}{$\rm N(1840)P_{11}$}\\
\hline
$J^P$& $\Delta\chi^2_{tot}$ &
\hspace*{-1mm}$\Delta\chi^2_{p\pi^0}$\hspace*{-1mm} &
\hspace*{-1mm}$\Delta\chi^2_{p\eta} $\hspace*{-1mm} &
\hspace*{-1mm}$\Delta\chi^2_{\KL} $\hspace*{-1mm} &
\hspace*{-1mm}$\Delta\chi^2_{\KS} $\hspace*{-1mm} \\
omitted & 565 & 315 & 46 & 248 & 176 \\
repl. by $1/2^-$ & 334 & 305 & 10 & 151 & 60  \\
repl. by $5/2^-$ & 211 & 236 & -6 & 64 & 158  \\
repl. by $7/2^-$ & 462 & 269 & 32 & 222 & 167  \\
repl. by $9/2^-$ & 291 & 208 & 17 & 158 & 32  \\
repl. by $3/2^+$ & 434 & 301 &  28 & 257  & 4   \\
repl. by $7/2^+$ & 516 & 305 &  45 & 177 & 155  \\
repl. by $9/2^+$ & 468 & 191 &  54 & 216 & 73 \\
\hline\hline
Resonance
&\multicolumn{5}{c}{$\rm N(1870)D_{13}$}\\
\hline
$J^P$& $\Delta\chi^2_{tot}$ &
\hspace*{-1mm}$\Delta\chi^2_{p\pi^0}$\hspace*{-1mm} &
\hspace*{-1mm}$\Delta\chi^2_{p\eta} $\hspace*{-1mm} &
\hspace*{-1mm}$\Delta\chi^2_{\KL} $\hspace*{-1mm} &
\hspace*{-1mm}$\Delta\chi^2_{\KS} $\hspace*{-1mm} \\
omitted & 320 & 204 &  30 & 63 & 103\\
repl. by $1/2^-$ & 274 & 77  & 53  & 16 & -21\\
repl. by $5/2^-$ & 129 & 97  & 11  & -19& 85    \\
repl. by $7/2^-$ & 215 & 144 & 19  &  21& 97    \\
repl. by $9/2^-$ & 162 & -29 & 37  &  28& 63    \\
repl. by $3/2^+$ & 262 & 43  & 57  & -1 & -13 \\
repl. by $7/2^+$ & 164 & 154 & 11  & 8  &  23 \\
repl. by $9/2^+$ & 204 & 87  & 35  & -36& 51  \\
\hline\hline
Resonance
&\multicolumn{5}{c}{$\rm N(2170)D_{13}$}\\
\hline
$J^P$& $\Delta\chi^2_{tot}$ &
\hspace*{-1mm}$\Delta\chi^2_{p\pi^0}$\hspace*{-1mm} &
\hspace*{-1mm}$\Delta\chi^2_{p\eta} $\hspace*{-1mm} &
\hspace*{-1mm}$\Delta\chi^2_{\KL} $\hspace*{-1mm} &
\hspace*{-1mm}$\Delta\chi^2_{\KS} $\hspace*{-1mm} \\
omitted & 311 & 161 & 6  & 368& 59\\
repl. by $1/2^-$ & 206 & 59  & ~~1 & 385 & -17 \\
repl. by $5/2^-$ & 39  & -77 & -11 & 305 &  30 \\
repl. by $7/2^-$ & 148 & -28 & -11 & 433 & 100 \\
repl. by $9/2^-$ & 87  & -52 &  0  & 292 & -1  \\
repl. by $1/2^+$ & 166 &  54 & -6  & 330 & 48 \\
repl. by $5/2^+$ & 149 & 136 & -27 & 368 & 31   \\
repl. by $7/2^+$ & 53  &  19 & -5  & 127 & 17  \\
repl. by $9/2^+$ & 133 &  94 & -15 & 299 & 2   \\
\hline\hline
\end{tabular}
\end{center}
\renewcommand{\arraystretch}{1.0}
\end{footnotesize}
\end{table}
With this state included a very
satisfactory description of all data sets (except for
$\KS$ polarisation, see below) was obtained up to
2100\,MeV. The $\chi^2$ was improved almost for all reactions. The fit
of the recoil polarisations without this state (where the effect is clearly
visible) is shown as dashed curves in
Figs.~\ref{fig:klrecoil},\ref{fig:ksrecoil}. The SAPHIR data on
$\rm K^0\Sigma^+$ production were added at the end of this analysis.
They are described with a $\chi^2=109$ for 120 data points.
The comparison of this solution with
resonances having other quantum numbers but a similar mass is made
in Table~\ref{kl_changes}.

\begin{figure}[b!]
\centerline{
\epsfig{file=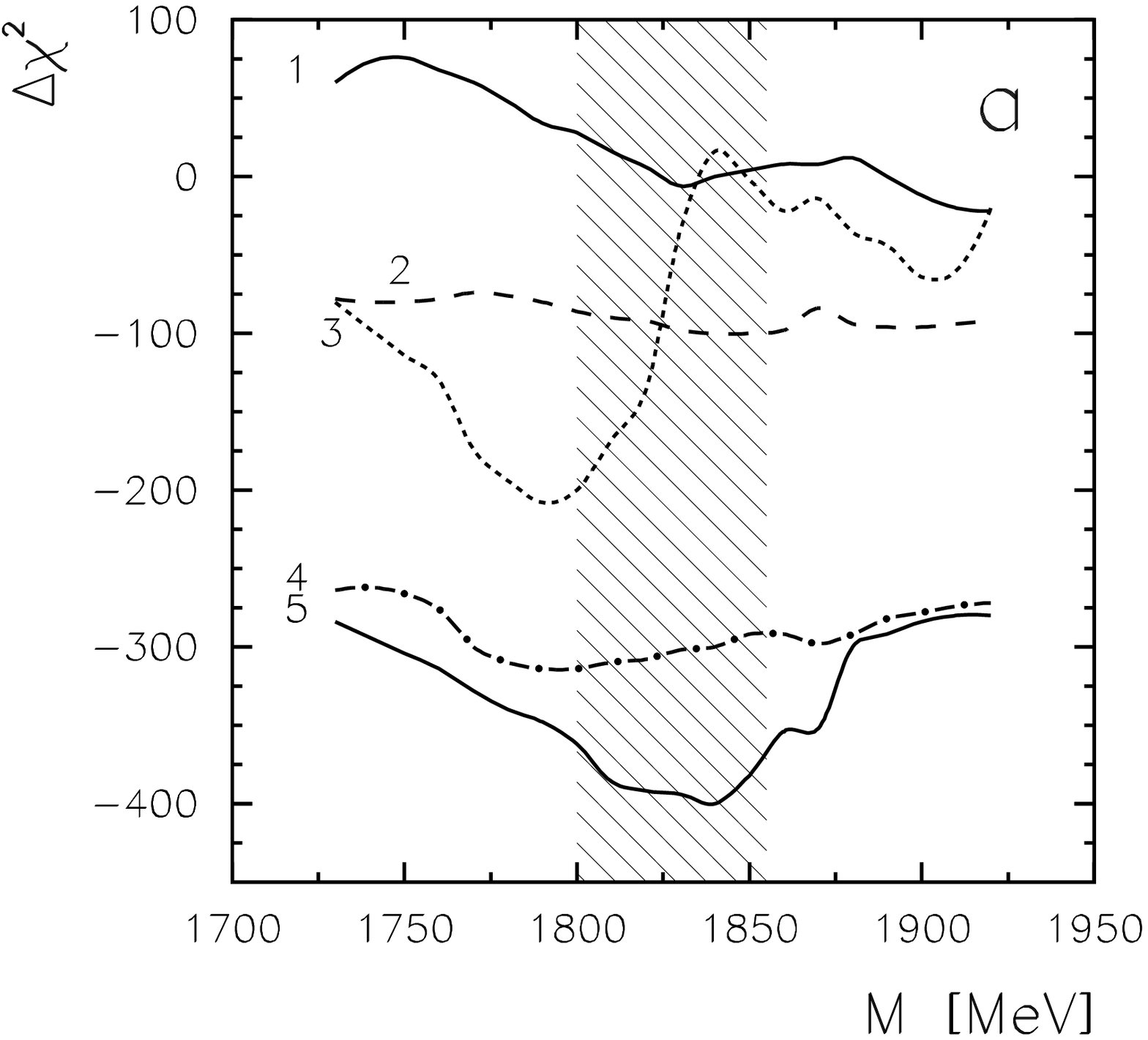,width=0.27\textwidth}\hspace*{-4mm}
\epsfig{file=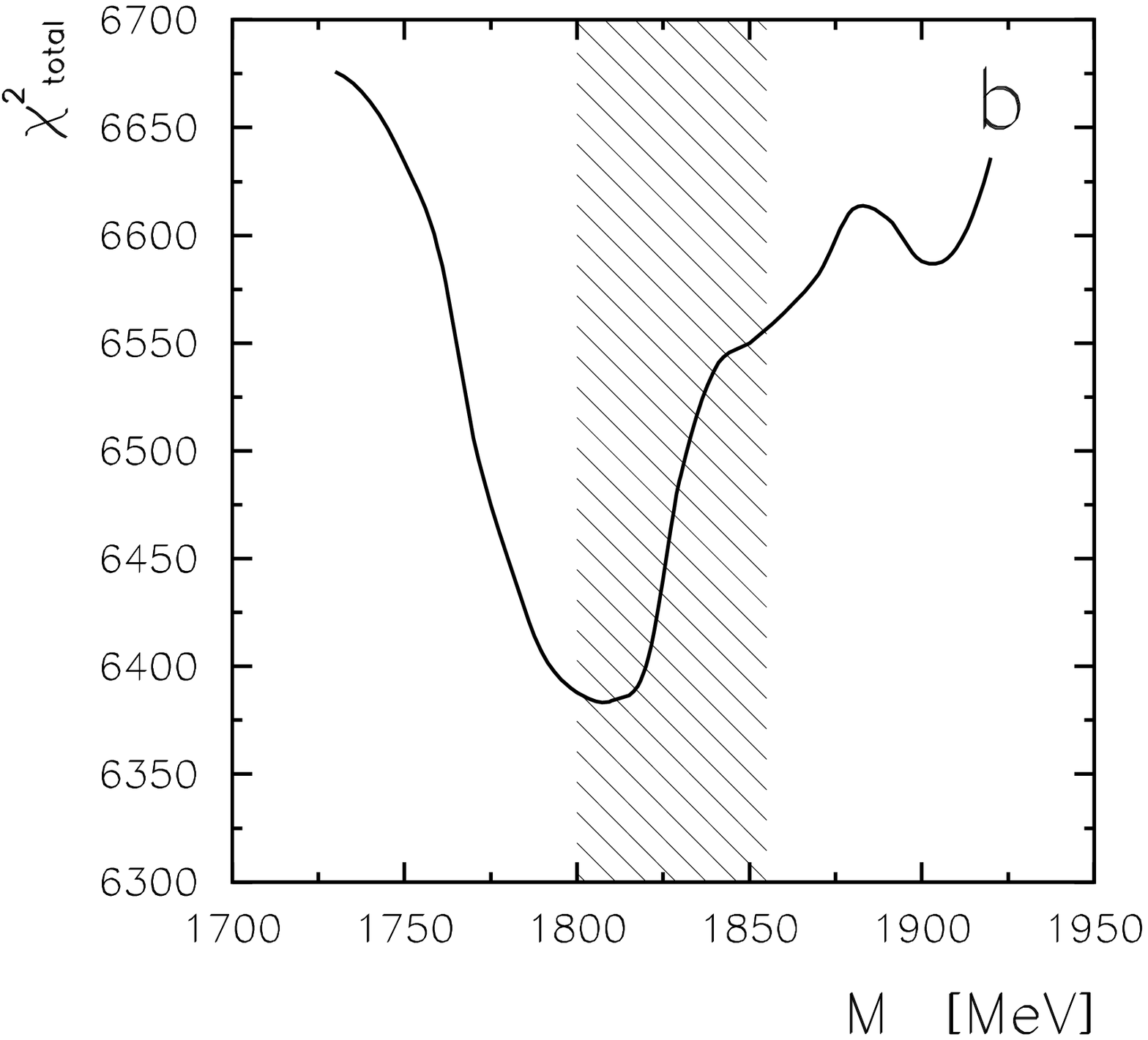,width=0.27\textwidth}}\vspace*{-4mm}
\centerline{
\epsfig{file=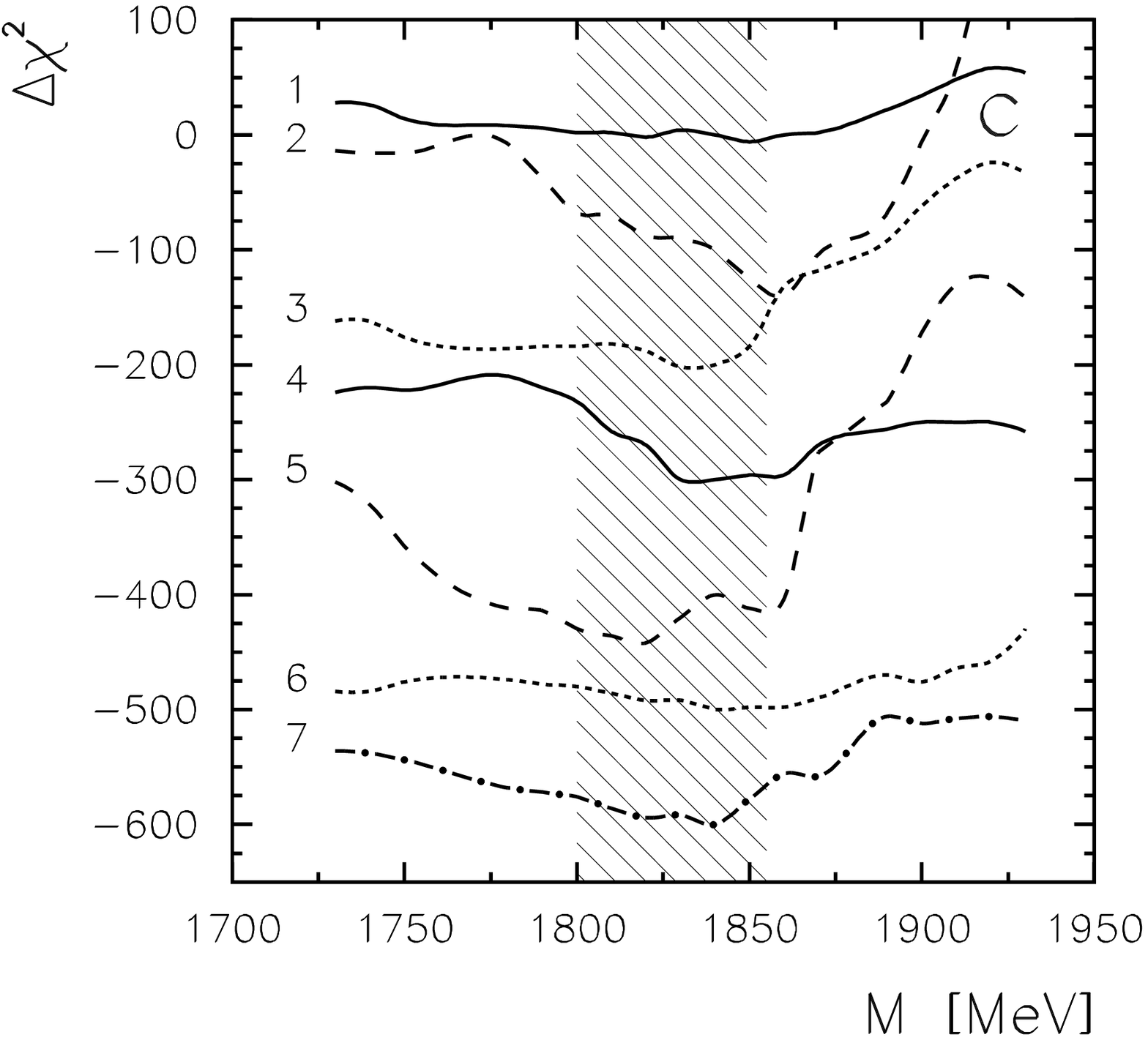,width=0.27\textwidth}\hspace*{-4mm}
\epsfig{file=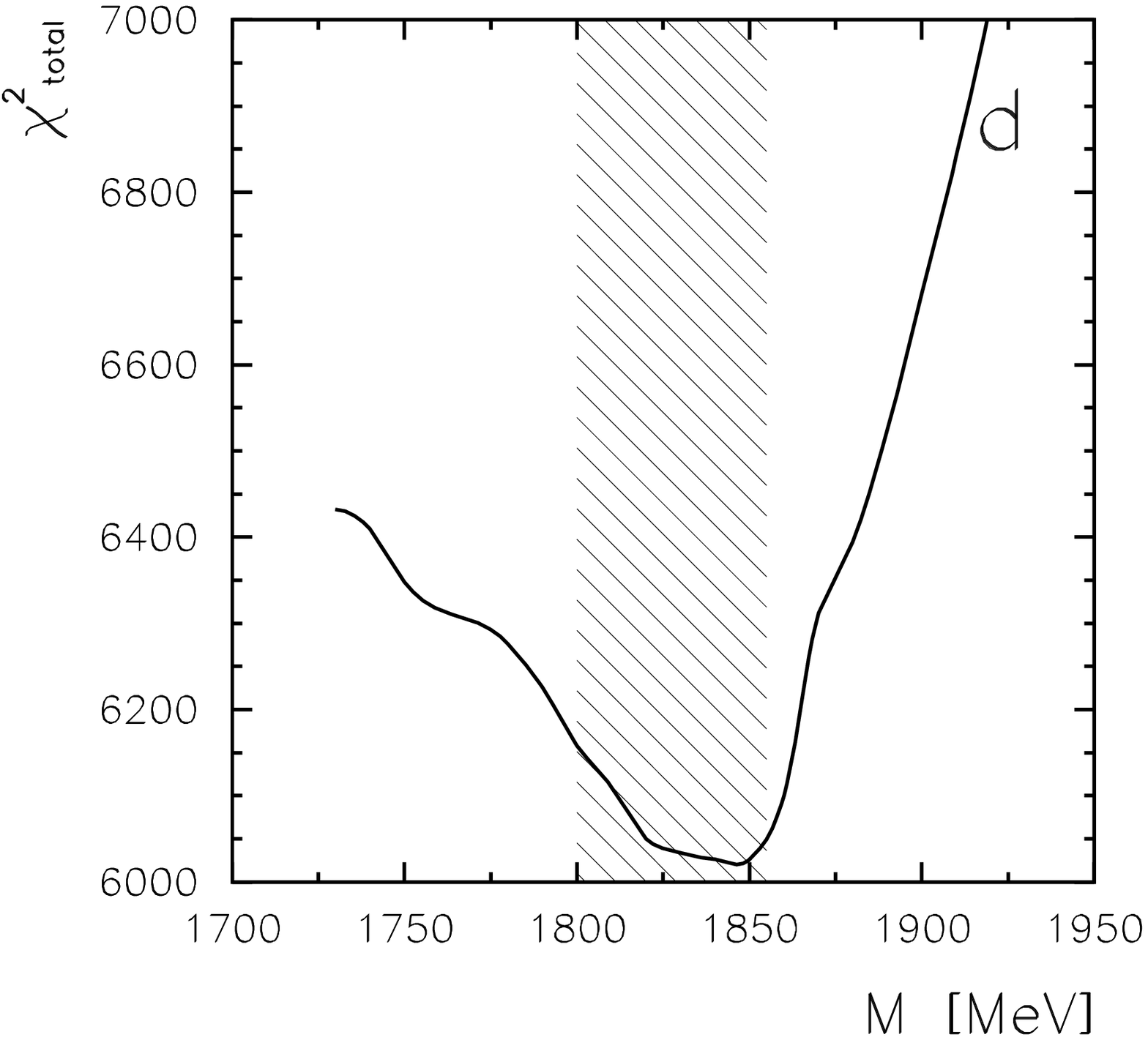,width=0.27\textwidth}\vspace*{-2mm}
}
\caption{The result of $\rm P_{11}(1840)$ mass scan:
a) 1 -- $\rm d\sigma/d\Omega$ for $\rm\gamma p\to \pi^0 N$ (CB-ELSA)
2 -- $\rm d\sigma/d\Omega$ for $\rm\gamma p\to \eta N$ (CB-ELSA)
3,4 -- beam asymmetry for $\pi^0$ and $\eta$ photoproduction (GRAAL),
5 -- $\rm d\sigma/d\Omega$ for $\pi^0$ photoproduction (GRAAL);
b) the sum of $\chi^2$ for the reactions shown in a);
c) 1 -- the SAPHIR $\KL$ $\rm d\sigma/d\Omega$, 2 -- the CLAS $\KL$
$\rm d\sigma/d\Omega$, 3 -- the $\KL$ recoil polarization (CLAS), 4 --
the SAPHIR $\KS$ $\rm d\sigma/d\Omega$, 5 -- the CLAS $\KS$
differential cross section, 6 -- the $\KS$ beam asymmetry, 7 -- the
$\KS$ recoil polarisation;
d) the total $\chi^2$ for all $\KL$ and $\KS$ reactions.}
\label{p11_1840}
\end{figure}

Fig.~\ref{p11_1840}\,a shows a mass scan of the resonance
($\chi^2$ as a function of the assumed  $\rm P_{11}$ mass).
In the scan, the mass of the  $\rm P_{11}$ mass was fixed 
to preset values while
all other fit parameters were allowed to adjust newly.
There is a clear $\chi^2$ minimum at 1810--1850\,MeV.
In Fig.~\ref{p11_1840}\,a,
the $\chi^2$ changes are shown for the fit of $\pi N$ and $\eta N$
differential cross sections and beam asymmetry data.
The $\rm P_{11}$ contribution leads to a minimum in $\chi^2$ for all
individual distributions, although the optimum is slightly lower for
the latest GRAAL beam asymmetry data.
The changes for the sum of $\chi^2$ for these reactions is shown in
Fig.~\ref{p11_1840}\,b. The correspondent pictures for the
reactions with $\KL$ and $\KS$ final states are shown in
Figs.~\ref{p11_1840}\,c,d. Here the $\chi^2$ distribution as well has a
minimum at 1840\,MeV almost for all reactions.
This fact provides strong evidence
that the $\rm N(1840)P_{11}$ is a genuine resonance and not an
artefact of some data. The width is determined to be 
$(140^{+35}_{-15})$\,MeV. This is the minimum value 
when the width is determined as a
function of the fitted mass: if the mass is shifted by 60\,MeV from the
central position the width increases almost by factor 2.
\par
\begin{figure}[b!]
\vspace*{-6mm}
\centerline{
\epsfig{file=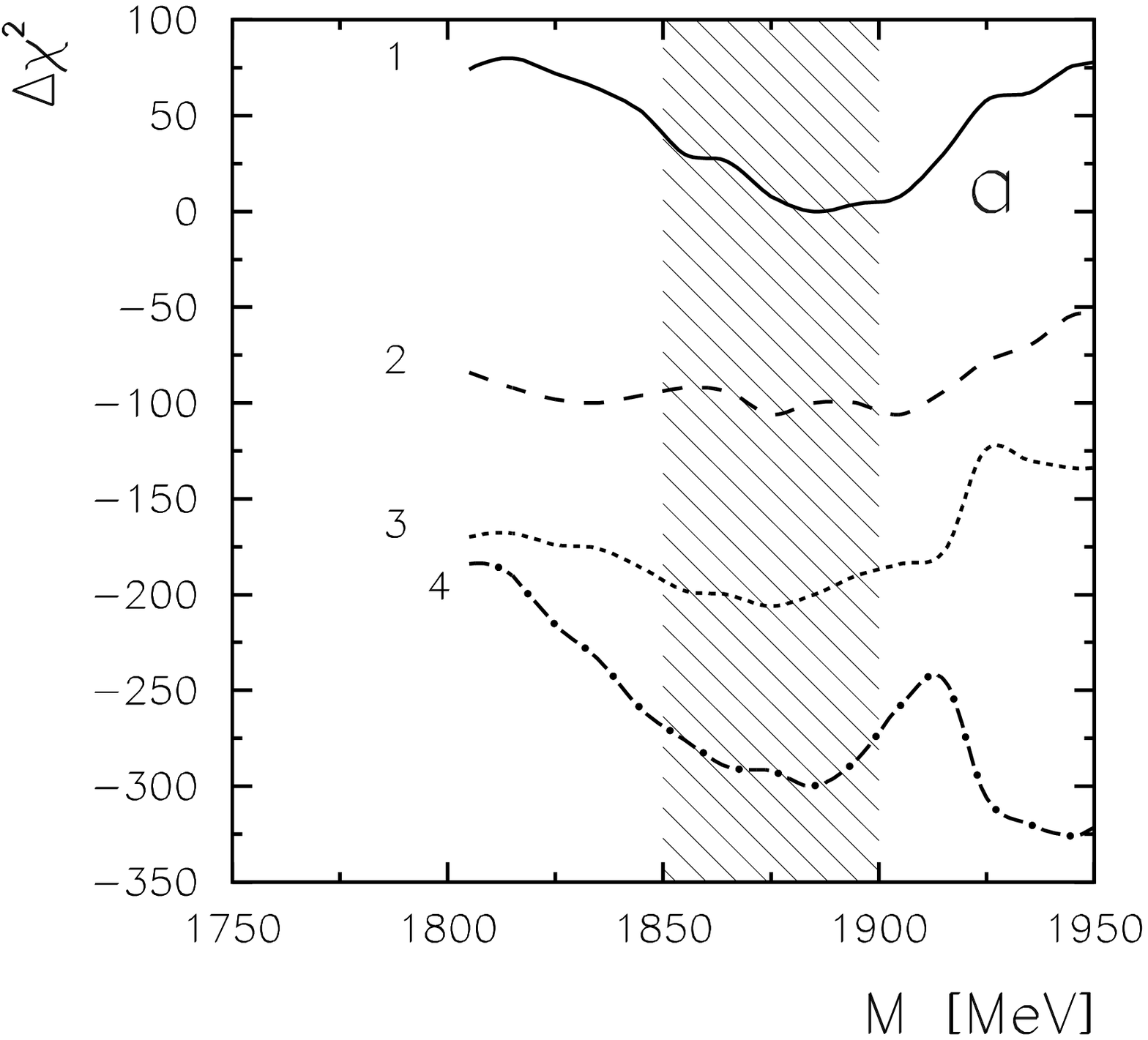,width=0.27\textwidth}\hspace*{-4mm}
\epsfig{file=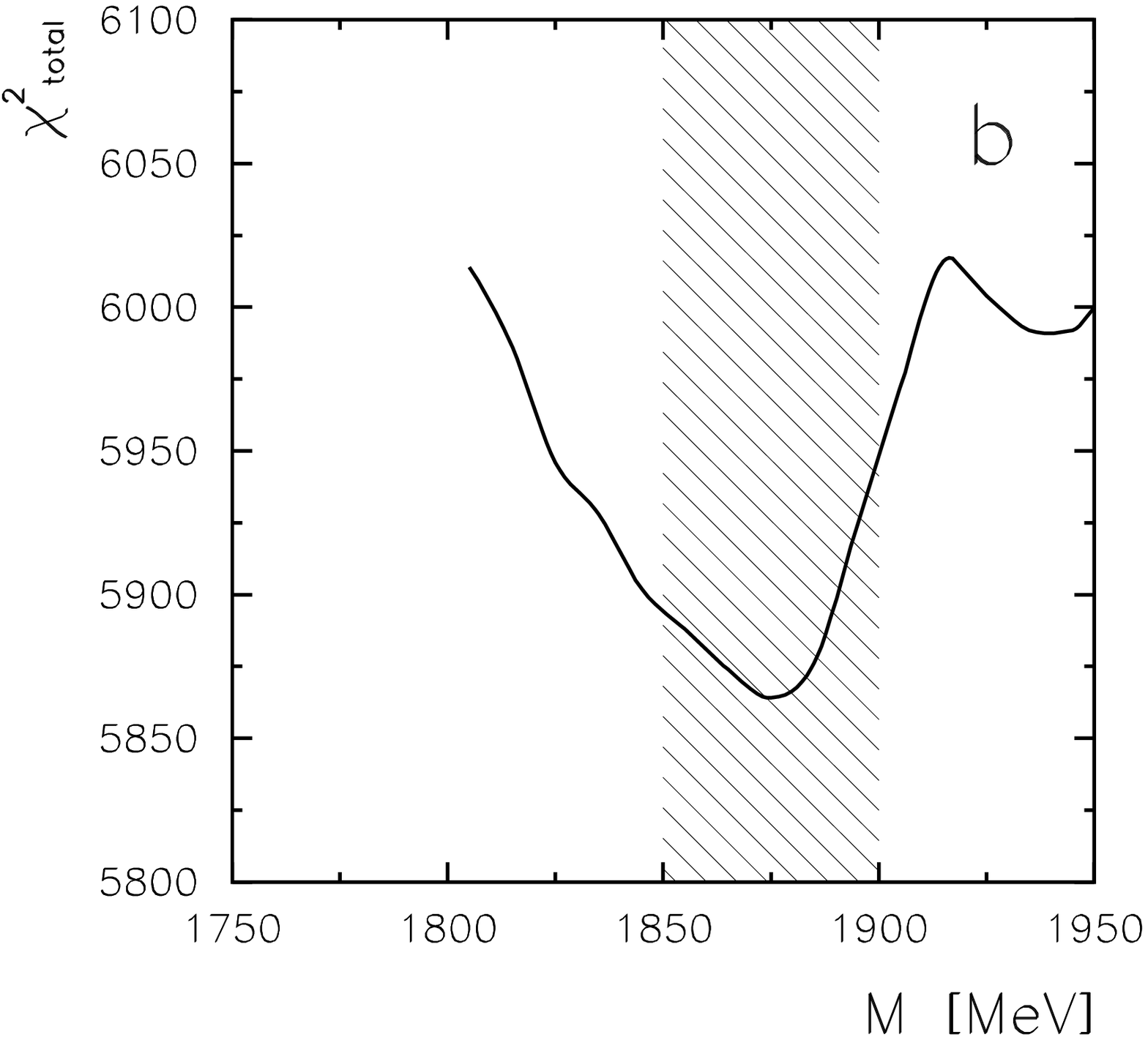,width=0.27\textwidth}\vspace*{-2mm}
}
\caption{The result of $\rm D_{13}(1870)$ mass scan:
a) 1 -- $\rm d\sigma/d\Omega$ for $\rm\gamma p\to \pi^0 N$ (GRAAL),
2 -- the $\KL$ recoil polarisation (CLAS),
3,4 -- $\rm d\sigma/d\Omega$ for $\rm\gamma p\to \KS$ reaction
(SAPHIR, CLAS);
b) the total $\chi^2$ for all $\KL$ and $\KS$ reactions.}
\label{fig:d13_1875}
\end{figure}
The  $\rm N(1840)P_{11}$ mass is considerably larger than that
of the PDG  $\rm N(1710)P_{11}$. This is a discrepancy but, possibly,
the $\rm N(1710)P_{11}$ could be split into two resonances,
an $\rm N(1670)P_{11}$ and an $\rm N(1840)P_{11}$.
An  $\rm N(1670)P_{11}$ is predicted as member of an
antidecuplet~\cite{Diakonov:2003jj}
and evidence was reported for a narrow resonance at 1670\,MeV
which might have $\rm P_{11}$ quantum numbers~\cite{Kuznetsov:2004gy}.
A mass scan was therefore performed searching for an additional
$\rm P_{11}$ state, fixing the mass of $\rm P_{11}(1840)$ at the
optimum value. The description of the CLAS differential cross sections
in both final channels was slightly improved if
this additional resonance had a large mass and a very wide width,
but there was no clear minimum in the $\chi^2$ distributions.
In the analysis of $\rm \pi N\to N\pi$ and $\rm N \pi\pi$ the $\rm P_{11}$
mass and width were determined to have values fully compatible with our
findings \cite{Manley}.

The recoil polarisation for $\KS$ production still had
systematic deviations between data and fit, and $\KL$ production
showed a discrepancy in the 2.2\,GeV mass region. Also the  recoil
polarisation for $\rm \gamma p\to\KL$ was not well described in the
2.2\,GeV mass region. As before resonances with different quantum
numbers were introduced one by one. Only adding contributions from
the $\rm D_{13}$ wave improved the picture. Two $\rm D_{13}$ states
needed to be introduced; the mass of the lower--mass state optimised
for  $(1875\pm 25)$\,MeV and a width of $(80\pm 20)$\,MeV.
The result of the mass scan for this state is shown in
Fig.~\ref{fig:d13_1875}. There is a clear minimum for the
GRAAL $\rm\pi N$ differential cross section
and for the $\KS$ (SAPHIR, CLAS) differential cross sections.
A rather shallow minimum is seen for the
$\rm\Lambda$ recoil polarisation.
The distribution of
sum of $\chi^2$
for all reactions with $\KL$ and $\KS$ final states is shown in
Fig.~\ref{fig:d13_1875}\,b.
The other state
was found at $(2166\pm35)$\,MeV, its width at $(280\pm 65)$\,MeV.
With this resonance, the recoil polarisation in $\rm \gamma p\to
\KL$ is described as well. Omitting this contribution from the fit
yields the fit which is shown in Figs.~\ref{fig:klrecoil}
and~\ref{fig:ksrecoil} as dashed line.
The mass scan of individual channels did not show a
clear minimum for this state. The $\chi^2$
for the recoil polarisation (differential cross section)
show asymmetric minima, shallow on the high--mass (low--mass)
side and steeper on the  low--mass (high--mass) side.
Only their sum gives a pronounced minimum in $\chi^2$.

\subsection{The high mass range ($>$\,2200\,MeV)}
At large photon energies, kaons are produced preferentially in forward
direction. The forward peaks are well reproduced by the fit which
assigns forward meson production to meson exchanges in the $t$--channel.
In the mass range covered by CLAS and SAPHIR,
the fractional contributions of the $\rm K$ and $\rm K^*$ exchanges
are 8\% and 22\% to the total cross section for $\KL$ and 25\% and
47\% for $\KS$, respectively. In the $\rm K^0\Sigma^+$ differential
cross section, there is no forward peak.  The $\rm K$ reggeized
exchange is supposed to be significantly suppressed, but the
fit does not find a significant contribution from $\rm K^*$ exchange
neither. 

In the $u$--channel, $\rm\Lambda$ and $\rm\Sigma$ exchanges
both contribute about 10\% to $\KL$ and to $\KS$.

\subsection{Discussion}
\subsubsection{Resonances with strong $\KL$, $\KS$ coupling}
For $\rm N(2000)F_{15}$ and $\rm N(1870)D_{13}$ the
masses were found to be about 70 MeV lower then in the previous
analysis of data on $\rm p\pi$ and $\rm p\eta$ final states only
\cite{Bartholomy:04,Crede:04}. As was pointed out in those
publications, these resonances only weekly contributed to $\rm p\pi$
and $\rm p\eta$ cross sections and were helpful to describe the
polarisation functions. In the present analysis we found that these
states have significant couplings to $\KS$ and adding the $\rm P_{11}(1840)$
state provided a flexibility to describe the polarisation
data. The $\rm F_{15}$ state has now mass and width which
are close to the values obtained by H\"ohler \cite{hoehler}. Moreover, the mass
and width of his solution can be used as fixed input values
without changing the quality of the description.

For $\KL$ production, the $\rm S_{11}$ partial wave provides $(48\pm 5)$\%
of the total cross section. The $\rm N(1720)P_{13}$ contribution
is the next strongest one with $(19\pm 4)$\%. The newly observed
states $\rm N(1840)P_{11}$ and $\rm N(2170)D_{13}$ contribute $3\pm 1$\% and
$(2\pm 0.8)$\%, repectively. For $\KS$ production, the $\rm S_{11}$ wave
provides the strongest resonance contributions, $(22\pm 6)$\%.
The second strongest resonance is $\rm D_{33}(1700)$ with
$12\pm 4$\% while $\rm N(1720)P_{13}$ contributes only about 1\% to the
cross section. The $\rm N(1840)P_{11}$ and $\rm N(2170)$ $\rm D_{13}$
resonances contribute on the level of
7\% and 1\% to the $\KS$ total cross section.

Although quite a large number of states
contribute only weakly to the total cross sections, their amplitudes are
important to describe the polarisation information. Here, the
interference between large and small amplitudes can significantly
change the polarisation function. The typical example is the beam
asymmetry in the $\eta$ photoproduction reaction where a small
contribution from $\rm N(1520)D_{13}$ changes  the behaviour of
the polarisation function dramatically.

\subsubsection{Four $\rm D_{13}$ resonances}
There are four nucleon resonances with quantum numbers
$I(J^P)=1/2(3/2)^-$ ($\rm D_{13}$). These quantum numbers can be formed with
intrinsic orbital angular momentum $L=1$ and intrinsic spin $S=1/2$,
or with $S=3/2$ and $L=1$ or $L=3$.
The lowest mass state at 1520\,MeV seems to be
dominantly in a ($J\!=\!3/2; L\!=\!1, S\!=\!1/2$)
state. Apart from $\rm\Delta(1232)P_{33}$, 
it provides the largest contribution to the $\rm p\pi^0$
photoproduction cross section. Its companion at 1700\,MeV, dominantly
($J\!=\!3/2; L\!=\!1, S\!=\!3/2$), makes a much smaller contribution. It seems
plausible that the two further states, 
$\rm N(1870)D_{13}$ and $\rm N(2170)D_{13}$, should have
a dominantly ($J\!=\!3/2; L\!=\!1, S\!=\!1/2$) configuration.
We note that the
spacings in mass square are
\be
\rm M^2(2166)-M^2(1875) = (1.19\pm 0.18) \ GeV^2\,,\\
\rm M^2(1875)-M^2(1520) = (1.20\pm 0.10) \ GeV^2\,,
\ee
which agrees with the N--Roper mass splitting
\be
\rm M^2(1440)-M^2(938)  = (1.19\pm 0.06) \ GeV^2\,.
\ee
This leads to the conjecture that $\rm N(1870)D_{13}$ and
$\rm N(2170)$ $\rm D_{13}$  could be the first and second radial excitation of
the $\rm N(1520)D_{13}$. It was shown in~\cite{Klempt:2002vp}
that all sequential
baryon resonances in a given partial wave have mass square spacings
of about $1.142\,\rm GeV^2$.
\par
\par
\subsubsection{Quark--model predictions for $\rm D_{13}$ resonances.}
Quark--model calculations predict a very large number of states.
This is exemplified here using the $\rm N\,D_{13}$ resonance series
and a comparison with the Bonn constituent--quark model~\cite{Loring:2001kx}.
The lowest mass states with negative parity have
$L=1$. The spatial wave function for
$L=1$ has mixed symmetry. Hence, the spin--flavour wave function
has to have mixed symmetry, it has to belong to a 70--plet. A mixed
symmetry spin--flavour wave function can be realised for $S=1/2$
and $S=3/2$. These two states can mix, thus forming the two states
belonging to the $1\,\hbar\omega$ band, $\rm N(1520)D_{13}$ and
$\rm N(1700)D_{13}$.
\par
In the third excitation band, the total orbital angular
momentum can be $L=1$ (with $S=1/2,\,3/2$) or $L=3$ (with $S=3/2$).
Spatial wave functions can be constructed now which are symmetric,
antisymmetric, or of mixed symmetry, resulting in a total of eight
different states. They do not only mix among themselves;
mixing with higher configurations belonging to the
fifth excitation band has to be considered as well.
As a result, the Bonn model predicts  $\rm N\,D_{13}$ resonances
at \underline{1472} and 1622\,MeV  in the first excitation band
and at 1918, \underline{1988}, 2146,
2170, 2190, 2223, \underline{2231}, and 2271\,MeV in the third
excitation band. (The quark model in~\cite{Capstick:1993kb} gives very
similar results.) The underlined
masses indicate those resonances which are dominantly spin 1/2
resonances within a 70--plet. It is conceivable that the photon couplings
of these resonances are larger than those of other states.
This conjecture could provide a natural explanation why
$\rm N(1870)D_{13}$ and  $\rm N(2170)D_{13}$ are observed and the
other states not. However,
model calculations of baryon decays do not reproduce this pattern.
If the other resonances exist in addition,
their discovery will require data of much higher
statistics and additional polarisation data. New data
from pion--induced reactions will likely be mandatory as well.
\par
\subsubsection{Do four $\rm S_{11}$ resonances exist\,?}
There are claims for four  nucleon resonances with quantum numbers
$\rm S_{11}$. A survey of these results can be found
in~\cite{Saghai:2004pt}. The observation of the
two high--mass states in photoproduction seems questionable
 since differential cross sections in the higher mass range are not
well reproduced; in pion--induced reactions, the introduction
of a third and fourth $\rm S_{11}$ resonance at
$(1846\pm 47)$ and $(2113\pm 70)$\,MeV improves data description
considerably~\cite{Chen:2002mn}.
In this analysis,
the two lowest mass states at 1535 and 1650\,MeV
are observed but we do not find any need for introducing
additional $\rm S_{11}$ states. The
pairs of resonances $\rm N(1535)S_{11}$
and $\rm N(1520)D_{13}$, $\rm N(1846)S_{11}$
and $\rm N(1870)D_{13}$, and $\rm N(2133)S_{11}$
and $\rm N(2170)D_{13}$, may provide reasonable spin doublets,
with small spin--orbit splittings. Resonances belonging to 
spin triplets (or degenerate quartets) like $\rm N(1650)S_{11}$
and $\rm N(1700)D_{13}$ are only weakly excited and their
radial excitations are not observed in the data discussed here.

\section{Summary}

Results of an analysis of hyperon photoproduction
in the  reactions $\rm\gamma p\to \KL$ and $\rm\gamma p\to \KS$
using data from the CLAS, LEPS and SAPHIR collaborations are presented.
The data are analysed in a combined fit with data on $\pi^0$ and
$\eta$ photoproduction. The SAPHIR and the CLAS data are compatible
only if a normalisation factor in the order of 0.85 is introduced. 
The combined
fit yields results which are compatible with the results on $\pi^0$
and $\eta$ photoproduction reported earlier by the CB--ELSA
collaboration but requires to introduce new baryon resonances. In
particular, a $\rm P_{11}$ state was observed in the region of
1840 MeV which contributes to almost all reactions.
The analysis highlights the existence of four $\rm D_{13}$
resonances, $\rm N(1520)D_{13}$, $\rm N(1700)D_{13}$,
$\rm N(1870)D_{13}$, and $\rm N(2170)D_{13}$.
A comparison with the Bonn quark model suggests
that the main component of their flavour wave functions
all belong to 70--plets; the weakly excited $\rm N(1700)D_{13}$
has dominantly intrinsic spin 3/2 while the other 3 resonances have mostly
spin 1/2.
\par

\subsection*{Acknowledgment}
We would like to thank the CB--ELSA/TAPS collaboration
for numerous discussions on topics related to this work.
We acknowledge
financial support from the Deutsche For\-schungs\-gemeinschaft
within the SFB/TR16. The
St. Peters\-burg group received funds from the
Russian Foundation for Basic Research (grant 04-02-17091).
U.~Thoma thanks for an Emmy Noether grant from the DFG.
A.~Ani\-sovich and A.~Sarantsev
acknowledge support from the Alexander von Humboldt Foundation.

\end{document}